\documentclass[twocolumn,aps,superscriptaddress,prl,amsmath]{revtex4-2}
\usepackage{txfonts}
\usepackage[latin9]{inputenc}
\usepackage{mathrsfs}
\usepackage{multirow}
\usepackage{amssymb}
\usepackage{graphicx,subfigure,float}
\usepackage{dcolumn}
\usepackage{bm}
\usepackage{color}
\usepackage[colorlinks, linkcolor=blue, citecolor=blue, urlcolor=blue]{hyperref}
\usepackage{lipsum}

\begin{document}
\title{General Stacking Theory for Altermagnetism in Bilayer Systems}
  \author{Baoru Pan}
\affiliation{Hunan Provincial Key laboratory of Thin Film Materials and Devices, School of Materials Science and Engineering, Xiangtan University, Xiangtan 411105, People's Republic of China}
   \affiliation{School of Physics and Optoelectronics, Xiangtan University, Xiangtan 411105, People's Republic of China}
  \author{Pan Zhou}
 \email{zhoupan71234@xtu.edu.cn}
\affiliation{Hunan Provincial Key laboratory of Thin Film Materials and Devices, School of Materials Science and Engineering, Xiangtan University, Xiangtan 411105, People's Republic of China}
\author{Pengbo Lyu}
\affiliation{Hunan Provincial Key laboratory of Thin Film Materials and Devices, School of Materials Science and Engineering, Xiangtan University, Xiangtan 411105, People's Republic of China}
 \author{Huaping Xiao}
    \email{hpxiao@xtu.edu.cn}
   \affiliation{School of Physics and Optoelectronics, Xiangtan University, Xiangtan 411105, People's Republic of China}
  \author{Xuejuan Yang}
   \affiliation{School of Physics and Optoelectronics, Xiangtan University, Xiangtan 411105, People's Republic of China}
\author{Lizhong Sun}
 \email{lzsun@xtu.edu.cn}
\affiliation{Hunan Provincial Key laboratory of Thin Film Materials and Devices, School of Materials Science and Engineering, Xiangtan University, Xiangtan 411105, People's Republic of China}
\begin{abstract}
Two-dimensional (2D) altermagnetism was recently proposed to be attainable in twisted antiferromagnetic bilayers providing an experimentally feasible approach to realize it in 2D materials. Nevertheless, a comprehensive understanding of the mechanism governing the appearance of altermagnetism in bilayer systems is still absent. In present letter, we address this gap by introducing a general stacking theory (GST) as a key condition for the emergence of altermagnetism in bilayer systems. The GST provides straightforward criteria to predict whether a bilayer demonstrates altermagnetic spin splitting, solely based on the layer groups of the composing monolayers. According to the GST, only seven point groups of bilayers facilitate the emergence of altermagnetism. It is revealed that, beyond the previously proposed antiferromagnetic twisted vdW stacking, altermagnetism can even emerge in bilayers formed through the symmetrically restricted direct stacking of two monolayers. By combining the GST and first-principles calculations, we present illustrative examples of bilayers demonstrating altermagnetism. Our work establishes a robust framework for designing diverse bilayer systems with altermagnetism, thereby opening up new avenues for both fundamental research and practical applications in this field.\\
\end{abstract}
\maketitle
\indent Recently, a novel magnetic phase known as altermagnetism has been proposed\cite{alter1, alter2, alter3, alter4, alter5, alter6}. In this phase, the total magnetic moment disappears, while the energy band dispersion exhibits spin-splitting along specific high-symmetry paths\cite{alter1}. This phenomenon holds significant promise for various areas within condensed matter physics\cite{spintronics1, spintronics2, spintronics3, spintronics4, spintronics5, caloritronics,multiferroics1, multiferroics2,superconductivity1, superconductivity2, superconductivity3, superconductivity4}. While altermagnetism in three-dimensional materials such as RuO$_2$\cite{RuO2_1,RuO2_2,RuO2_3,spintronics6,spintronics7} and CrSb\cite{alter1} has been theoretically proposed and experimentally confirmed, reports on altermagnetism in two-dimensional (2D) materials remain scarce. A recent study suggests that twisted antiferromagnetic bilayers are promising candidates for realizing altermagnetism in 2D systems\cite{tang_liming}. The proposal, involving van der Waals (vdW) stacking of 2D materials, provides an experimentally feasible approach to realize altermagnetism in 2D materials. However, the comprehension of the symmetrical connection between the composing monolayers and the resulting altermagnetic bilayer is still limited. Furthermore, in addition to twisted vdW stacking of bilayers, are there any other methods such as direct stacking available for constructing altermagnetic bilayers? A comprehensive understanding of the mechanisms underlying the occurrence of altermagnetism in bilayer systems is crucial for advancing both theoretical and experimental research in the field.\\
\begin{figure}
	\center
	\includegraphics[trim={0.0in 0.0in 0.0in 0.0in},clip,width=3.4in]{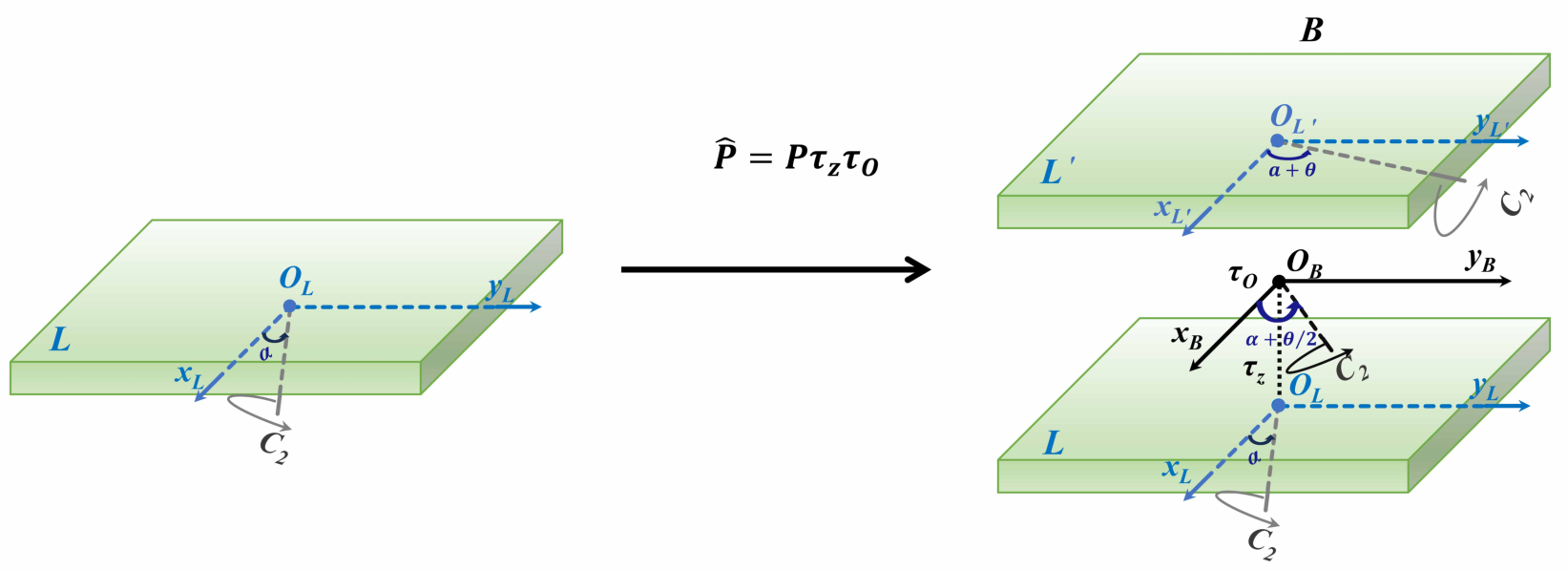}
	\caption{ Schematic of the construction of a bilayer from two monolayers. Applying a stacking operator $\hat{P}$, $L'$ can be created from the initial monolayer $L$, and both $L$ and $L'$ together constitute the bilayer, denoted as $B$. After the construction of the bilayer, a unified coordinate system, denoted as $x_{B}O_{B}y_{B}$, is employed to describe the operation matrices.}\label{fig1}
\end{figure}
\begin{figure*}
	\center
	\includegraphics[trim={0.0in 0.0in 0.0in 0.0in},clip,width=6.0in]{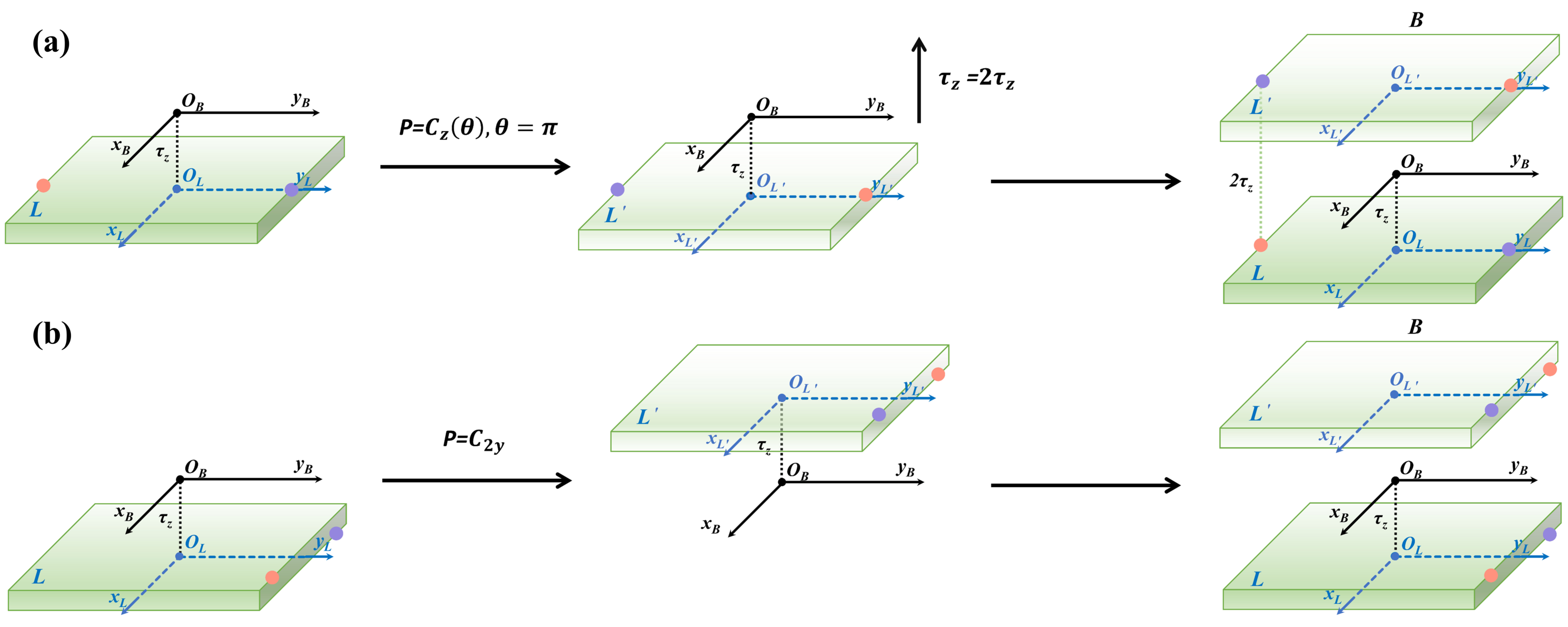}
	\caption{Schematic of construction of bilayer from the initial monolayer with the stacking operator of (a) ${C}_{z}(\theta),\theta = \pi$ and (b) $C_{2y}$. The purple and orange solid dots represent atoms in the unit cell that change under corresponding symmetry operation.}\label{fig2}
\end{figure*}
\indent In this Letter, we propose a general stacking theory (GST) for the emergence of altermagnetism in bilayer systems by applying group theory analysis across all 80 layer groups. The GST provides straightforward criteria to predict whether a bilayer demonstrates altermagnetic spin splitting, solely based on the layer group of the composing monolayers. Moreover, the GST also provides valuable insights into the necessary symmetry operations on the composing monolayers for the consistent emergence of altermagnetism in a bilayer. Through the integration of GST and first-principles calculations, we present two illustrative examples demonstrating altermagnetism. Inspired by the similar symmetry prerequisites for both altermagnetism and piezoelectricity, we unveil nontrivial piezoelectric properties in the seven types of altermagnetic bilayers. The compatibility of both properties suggests that the properties of the bilayers can be modulated by an external electric field via field-induced strain and structural distortions.\\
\indent The bilayer ($B$) can be directly constructed from a magnetic monolayer $L$ and its counterpart $L'$ generated from $L$ using the stacking operator ($\hat{P}$), as illustrated in Fig. \ref{fig1}. Note that we assume the two monolayers, $L$ and $L'$, exhibit interlayer antiferromagnetic coupling (A type antiferromagnetic coupling). To obtain an altermagnetic bilayer, we need to understand the relationship between the symmetry groups of monolayer $L$ (point group $G_L$ and layer group $LG_L$) and bilayer $B$ (point group $G_B$ and layer group $LG_B$) generated through the stacking operator $\hat{P}$. The $\hat{P}$ can be defined as $\hat{P}$=$P$$\tau_z$$\tau_O$, where the terms of $P$, $\tau_z$, and $\tau_O$ are point group operations, translation along the $z$-direction and translation in the $x_BO_By_B$ plane, respectively. However, the coordinate origin point of a monolayer (within the monolayer) and a bilayer (between the two monolayers) are generally different. Therefore, the stacking operator $\hat{P}$ matrices are hardly determined from the symmetry operation matrices of the initial monolayer and generated bilayer. To address this issue, we adopt the coordinate system of the bilayer, with the monolayers $L$ and $L'$  symmetrically positioned below and above the plane of $k_z = 0$, respectively. Under such condition, all symmetry operations of the monolayer $L$ can be classified into two distinct types, depending on whether they transform $z$ into -$z$: $\left \{{Q}^{+}\right \}=\left \{E,{C}_{z}(\theta),{M}_{\beta }\right \}$ and $\left \{{Q}^{-}\right \}=\left \{{C}_{2\alpha},{M}_{z},I,{S}_{nz}\right \}$. While ${Q}^{+}$ are still directly feasible symmetry operations on the monolayer, the use of ${Q}^{-}$ requires a $2\tau_z$ translation along the -$z$ direction. Consequently, the initial 3 $\times$ 3 matrix representations of the symmetry operations of monolayers need to be revised into 4 $\times$ 4 matrices, as shown in Tab. S1 in the Supplemental Material (SM)\cite{sup}. This revision provides us an effective approach for generating the stacking operator $\hat{P}$ when considering vdW stacking bilayers restricted by the twisted and point group operations of the monolayer. The matrix representations of the stacking operator $\hat{P}$ for the seven categories are presented in Tab. S1 of the SM\cite{sup}. Detailed discussions on this topic can be found in Sec. I of the SM\cite{sup}. A bilayer can be constructed when a stacking operator $\hat{P}$ acts on a monolayer. The matrix representations of $\hat{P}$ [$D_{\hat{P}}(\hat{Q})$] can also be divided into two major categories according to ${Q}^{+}$ and ${Q}^{-}$. The stacking operator $\hat{P}$ derived from ${Q}^{+}$ includes a $2\tau_z$ translation along the z-direction, as shown in Fig. \ref{fig2}(a). In contrast, the stacking operator $\hat{P}$ derived from ${Q}^{-}$ does not require such translation. Moreover, the stacking operator $\hat{P}$ can also include specific in-plane fractional translations $\tau_O$ depending on the point group symmetry of the monolayer $L$. Further details about the stacking operator $\hat{P}$ and their matrices including interlayer fractional translations can be found in Tabs. S1 and S2 of the SM\cite{sup}.\\
\indent To construct a bilayer, it is essential to determine the operations of the stacking operator $\hat{P}$ for a monolayer $L$ with specific point group. To encompass all potential bilayer configurations, we can categorize the operations corresponding to $\hat{P}$ into two groups: (i) The operations that belong to conventional crystallographic point group operations. The bilayer formed through this type of $\hat{P}$ corresponds to a direct stacking of individual monolayer $L$ and symmetrically restricted $L'$. The distinctive characteristic of the bilayer obtained in this way is that the two in-plane primitive vectors of the bilayer exactly match those of the initial monolayer $L$. Namely, the in-plane components of the 2D lattice of both the initial monolayer and the final bilayer remain unchanged under the stacking operator $\hat{P}$. If the point group of the monolayer $L$ is ${G}_{L}$ and its corresponding symmetry point group $G$ of 2D Bravais lattice belong to one of the four point groups of $C_{2h}$, $D_{2h}$, $D_{4h}$, and $D_{6h}$, the operations of $\hat{P}$ must be the element of the factor group ${G}_{P} = G/{G}_{L}$; (ii) operations with certain twist angle will construct bilayer through combination of two twisted monolayers $L$ and $L'$. The number of the operations of this type of $\hat{P}$ depends on the rotation angle $\theta$ around the z-axis [${C_z(\theta)}$] that produces the bilayer as a 2D crystal. It should be noted that for the twisted bilayer, the operations $\hat{P}$ must incorporate an additional translation along the $z$ axis. Furthermore, the in-plane unit cell of the resulting bilayer is typically expanded in comparison with that of the original monolayer.\\
\indent As mentioned above, when a bilayer is generated from initial monolayers, the point group of the monolayer and corresponding bilayer should be connected. Assuming a bilayer contains monolayer $L$ and $L'$, we can categorize the symmetry operation elements of the bilayer into two groups. The first group consists of the shared symmetry elements of $L$ and $L'$, which we refer to as the intersection set. This set can be represented as $O_i={G}_{L}\cap {G}_{L'}$, where ${G}_{L}$ is the initial point group of $L$ and ${G}_{L'}$ is obtained by applying the stacking operator $\hat{P}$ to ${G}_{L}$: ${G}_{L'}=\hat{P}{G}_{L}\hat{P}^{-1}$. The second group corresponds to the symmetry operations that interchange the two layers, transforming $L$ into $L'$ and vice versa. These elements are defined by the intersection of the cosets $\hat{P}{G}_{L}$ and ${G}_{L}\hat{P}^{-1}$ as $O_e=\hat{P}{G}_{L} \cap {G}_{L}\hat{P}^{-1}$. If $O_e$ is not empty, we refer to these elements as the exchange set. Consequently, the symmetry group of the bilayer, denoted as $G_{B}$, is a combination of the intersection set $O_i$ and the exchange set $O_e$, namely $G_{B}=O_i \cup O_e$. Therefore, the point group of a bilayer can be determined by considering the point group of the monolayer and the stacking operator. Similarly, the layer group of a bilayer can be obtained through the same process by considering the layer group of the monolayer.\\
\indent To comprehend the symmetry condition necessary for the emergence of altermagnetism in the vdW stacking bilayers, it is essential to revisit the spin group theory for altermagnetic phases in nonrelativistic collinear systems, as originally proposed by Libor and his colleagues\cite{alter1,alter2}. Different from conventional ferromagnetism and antiferromagnetism, the nontrivial spin space groups of altermagnetic phase can be written as ${\textbf{R}}_{s}^{III}=[E||\textbf{H}]+[{C}_{2}||\textbf{G}-\textbf{H}]$. The term $\textbf{H}$ denotes a halving subgroup of $\textbf{G}$, where the symmetries establish connections between the same-spin sublattices in real space. On the other hand, the symmetries in the cosets $\textbf{G}-\textbf{H}$ connect the opposite-spin sublattices in real space. To facilitate the emergence of altermagnetic phase, the $\textbf{G}-\textbf{H}$ should not include space inversion $\mathcal{I}$ and horizonal mirror operation $M_z$ in their rotation part\cite{spin_space_1, spin_space_2, spin_space_3, alter1}. These operations will lead to the degeneracy of opposite spin bands across the whole 2D reciprocal space. In other words, if $M_z$ and the spin reversal operation $U$ as defined in the reference \cite{hidden_spin} or the space inversion operation $\mathcal{I}$ and the time-reversal operator $\mathcal{T}$ remain in the antiferromagnetic coupling bilayers, the spin polarization is hidden at all k-points. The $\mathcal{T}$ that reverses both the wave vector $\textbf{k}$ and the spin $s$ leads to $\mathcal{T}\epsilon(s,\textbf{k})=\epsilon(-s,-\textbf{k})$; the spatial inversion operation $\mathcal{I}$ that reverses $\textbf{k}$ while keeping $s$ invariant results in $\mathcal{I}\epsilon(s,\textbf{k})=\epsilon(s,-\textbf{k})$; the combined $\mathcal{I}\mathcal{T}$ operation will also force $\mathcal{I}\mathcal{T}\epsilon(s,\textbf{k})=\epsilon(-s,\textbf{k})$ and a compound operation of mirror reflection $M_{z}$ and $U$ would force $M_{z}U\epsilon(s,\textbf{k})=\epsilon(-s,\textbf{k})$. Through an analysis of all potential point group symmetry operations, we determined that only two types of operations, ${C}_{2\alpha}$ and ${S}_{4z}$, are capable of connecting the sublattices with opposite spins and guaranteeing the existence of altermagnetism in bilayers. Based on these prerequisites of the symmetric operations, we have analyzed all 2D point groups of bilayer systems and identified seven point groups of bilayers that meet the above criteria: C$_{2}$, D$_{2}$, D$_{3}$, D$_{4}$, D$_{6}$, D$_{2d}$, and S$_{4}$. The common symmetry operation of the first six point groups is in-plane $C_2$ rotation.\\
\indent Except pure point group operations, we also consider potential fractional translations within the layer groups that describe the symmetries of 2D materials. Using all symmetry operations of monolayers and the transformation operation $\hat{P}$, we identify all monolayer layer groups that can be transformed into the seven targeted bilayer point groups, and the detailed results can be found in Tab. S2 of the SM\cite{sup}. With above analyses, a GST for altermagnetism in bilayers can be established. The GST allows us to determine whether a bilayer exhibits altermagnetic spin splitting solely based on the layer group of the composing monolayer. The GST can be briefly expressed as follows: (i) altermagnetism in bilayers is only possible when the layer group ${G}_{L}$ of the initial monolayer falls within one of the following categories: $L_{8-22}$, $L_{27-54}$, $L_{57-76}$, or $L_{78-80}$ (indicated by $\checkmark$ in Tab. S2 in the SM\cite{sup}); (ii) with few exceptions (stacking operators $\hat{P} = {C}_{z}(\theta)$ as listed in Tab. S1), the point groups of a twisted bilayer with any twisted angles reserving 2D crystalline are identical resulting in similar altermagnetic properties, such as layer groups of 27-36, which is feasible for realizing altermagnets in experiments; (iii) if the initial monolayer $L$ exhibits an in-plane $C_2$ rotation, the twisted bilayer derived from it with a twist angle of  $\theta$ must possess an in-plane $C_2$ axis. The angle between the $C_2$ axis of initial monolayer and final bilayer is $\theta/2$ (see Fig. \ref{fig1}), which generally maintains altermagnetism in the bilayer. In addition, even if the transformation operation is the identity, altermagnetic spin splitting can still occur provided that the bilayer point group belongs to one of the aforementioned seven point groups. It is worth noting that the choice of the origin for some layer groups requires attention in constructing altermagnetic bilayer, and the details can be found in the Sec. IX of the SM\cite{sup}.\\
\indent As a further stage, we apply the GST to analyze the antiferromagnetic twisted vdW stacking bilayers with altermagnetism previously reported by Tang et al.\cite{tang_liming}. To construct a twisted crystalline bilayer, the Bravais lattices of 2D materials must be rectangular, square, or hexagonal lattice\cite{twisted_bilayers}. It is worth noting that the square and hexagonal lattice can be considered as special cases of rectangular lattices. For twisted vdW stacking bilayers, the aspect ratio $\rho$ of the initial monolayer lattice must be the square root of a rational number $\sqrt{p{/}q}$ to form a crystal, where $p$ and $q$ are positive integers\cite{twisted_bilayers}. The monolayers constituting the previously proposed altermagnetic bilayers include hexagonal 1T-NiCl$_{2}$, 2H-NiCl$_{2}$, CrI$_{3}$, and square CrN\cite{tang_liming}. Square and hexagonal lattices meet this requirement since their aspect ratios are 1 and $\sqrt{3}$, respectively. Therefore, the lattices can be twisted at specific angles to match with each other to form crystals. The layer groups (point groups) of the monolayers of these materials are $L_{72}$ $(D_{3d})$, $L_{78}$ $(D_{3h})$, $L_{71}$ $(D_{3d})$, and $L_{61}$ $(D_{4h})$ for 1T-NiCl$_{2}$, 2H-NiCl$_{2}$, CrI$_{3}$, and CrN, respectively. All the point groups include an in-plane $C_2$ operation. Twisted bilayers constituted by them must also possess in-plane $C_2$ that connects the altermagnetic bilayer following to GST (iii). Interestingly, we find that the point groups of the four bilayers ($D_{3}$, $D_{3}$, $D_{3}$, and $D_{4}$ for bilayer 1T-NiCl$_{2}$, 2H-NiCl$_{2}$, CrI$_{3}$, and CrN, respectively) remain unchanged under any twisted angle $\theta$, as shown in Tab. S2. Namely, in addition to the typical rotation angles reported previously\cite{tang_liming}, the four bilayers will exhibit altermagnetism under any permissible twisted angles according to the GST. To demonstrate that a twisted bilayer with a rectangular lattice can also achieve an altermagnetic phase, we conducted further examination on a twisted bilayer CrBrSe, depicted in Fig. S2. Detailed symmetry analysis and energy band structure confirm that it is indeed an altermagnetic 2D material. The detailed discussions are provided in the Secs. III-V of the SM\cite{sup}.\\
\begin{figure}
	\center
	\includegraphics[trim={0.0in 0.0in 0.0in 0.0in},clip,width=3.5in]{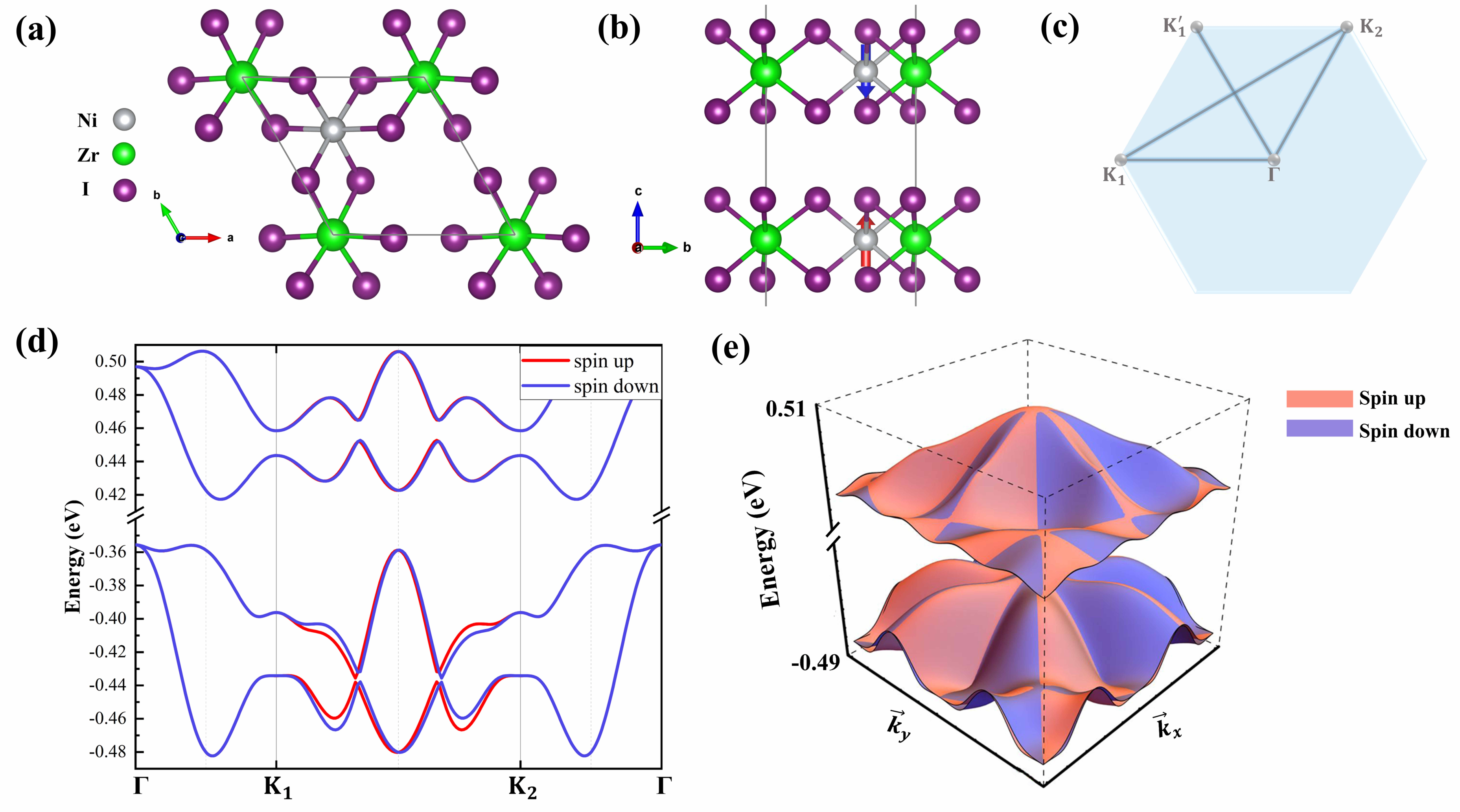}
	\caption{(a) Top and (b) side views of the bilayer NiZrI$_{6}$ obtained with $\hat{P}=E$. The grey, green and dark purple spheres represent Ni, Zr, and I atoms, respectively. Red and blue arrows indicate the local magnetic moments of magnetic ions. (c) First Brillouin zone, (d) band structure, and (e) three-dimensional band structure of bilayer NiZrI$_{6}$.
}\label{fig3}
\end{figure}
\indent The GST suggests that altermagnetism can also emerge in seven types of antiferromagnetic vdW stacking bilayers formed through symmetrically restricted direct stacking of individual monolayers without twisting. The point groups of the resulting bilayers can be C$_2$, D$_2$, D$_3$, D$_4$, D$_6$, D$_{2d}$, and S$_4$, respectively. In this scenario, the unit-cell of the bilayer is minimized and the first Brillouin zone is maximized, leading to an increased spin splitting size in reciprocal space. Moreover, the relatively small unit-cell of the bilayer makes it more efficient for the theoretical investigations based on first-principles method. We take layer group $L_{67}$ with a point group of $D_{3}$ as an example to illustrate this scenario. According to Tab. S2 and Fig. S5, the point group of twisted bilayer constituting monolayers with the layer group is either $C_{3i}$ or $C_{3h}$, which cannot sustain the altermagnetic phase. Nevertheless, if the $L'$ layer is created through each of the four stacking operators of $\hat{P}=E, C_{z}(\pi), \left \{ C_{z}(\pi)\mid \frac{1}{3},\frac{2}{3}\right \}$, or $\left \{ C_{z}(\pi)\mid \frac{2}{3},\frac{1}{3}\right \}$ as depicted in Fig. S5, an extra in-plane $C_2$ symmetry is introduced into the bilayer with a point group of $D_{3}$, thereby permitting the existence of altermagnetism in accordance with the GST. By searching the 2D material database C2DB\cite{c2db}, we find that the layer group and point group of 2D ferromagnetic material NiZrI$_{6}$ are $L_{67}$ and $D_{3}$, respectively. The top and side views of the lattice structures of the bilayer NiZrI$_{6}$ obtained through stacking operator $\hat{P}$ = $E$ are illustrated in Figs. \ref{fig3}(a) and \ref{fig3}(b), respectively. The discussions about its stability can be found in the Sec. III of the SM\cite{sup}. The band structure of the bilayer NiZrI$_{6}$, as shown in Fig. \ref{fig3}(d), exhibits typical altermagnetic characteristics. The spin momentum locking of the three-dimensional band structure under the collinear antiparallel spin configuration, as shown in Fig. \ref{fig3}(e), aligns with the nontrivial spin group of  [E$||$\textbf{C$_3$}]+[C$_{2}||$\textbf{D$_{3}$}-\textbf{C$_{3}$}]\cite{alter1}. These operations lead to the presence of three spin-degenerate nodal lines. Theoretically, if an antiferromagnetic coupled bilayer satisfies the symmetry requirements of the altermagnetic phase, its ground state is possible to be altermagnetic. However, exhibiting the altermagnetic characteristics, namely alternate distribution of spin splitting in reciprocal space, also depends on the anisotropy of the electronic crystal potential for opposite spin magnetic atoms\cite{alter1}. As for a bilayer, the anisotropy can be divided into two aspects: intra-layer anisotropy and inter-layer anisotropy, both of them can induce the spin splitting. The spin splitting of the NiZrI$_{6}$ bilayer is mainly determined by the strength of interlayer interactions, or inter-layer anisotropic electronic crystal potential for opposite spin magnetic atoms. The evolution of the spin splitting with the layer distance of the bilayer is shown in Fig. S9. The spin splitting of NiZrI$_{6}$ bilayer will disappear when the interlayer distance is enlarged. The detailed discussions can be found in the Sec. VI of the SM\cite{sup}. In addition, from Figs. S10(a)-S10(c), we can see that when the stacking operator contains certain interlayer fractional translations, such as $\left \{ E\mid \frac{1}{2},0\right \}$ and $\left \{ E\mid \frac{1}{2},\frac{1}{2}\right \}$, the point groups of the bilayers transform into C$_{2}$ that belongs to the altermagnetic bilayer point groups according to the GST. A detailed discussion of this part can be found in the Sec. VII of the SM\cite{sup}. Except NiZrI$_{6}$, through comprehensive exploration of magnetic monolayers and analysis of their layer groups in C2DB, we have identified a total of 392 2D materials with 22 layer groups that can potentially form altermagnetic bilayers. Details of the representative materials are provided in the Sec. VIII of the SM\cite{sup}.\\
\indent In summary, we present a GST for bilayers with altermagnetism. The GST offers effective guidance on how to design altermagnetic bilayers solely based on layer group of initial monolayer and stacking operators. Certainly, it should be emphasized that the GST results from the antiferromagnetic coupling between the two monolayers. Antiferromagnetic coupling between the layers is the precondition to apply the GST. If the magnetic ground states of certain bilayers do not show antiferromagnetic coupling, several experimental techniques can be employed to adjust the interlayer magnetic coupling into antiferromagnetic one. The techniques include modifying the stacking order \cite{tunable_mag1, tunable_mag2}, electrostatic doping \cite{tunable_mag4}, or applying an external electric field\cite{tunable_mag5,tunable_mag6}. Then the GST can be used to determine whether the bilayer is an altermagnetic material or not. One common feature for altermagnetic bilayer predicted by GST is the lack of spatial inversion symmetry. Such feature aligns with the prerequisite for bilayers exhibiting piezoelectricity. Notably, piezoelectricity is expected to coexist in all the seven types of altermagnetic bilayers predicted by GST, as elaborated in the Sec. X of the SM\cite{sup}. Such stacking dependent altermagnetic-ferroelectric states offer a novel avenue for manipulating spin polarization in spintronics and multi-state storage applications.\\
\indent This work is supported by the Excellent Youth Funding of Hunan Provincial Education Department (22B0175), Natural Science Foundation of Hunan Province (No. 2023JJ30572), the Scientific Research Fund of Hunan Provincial Education Department (18A051), the National Natural Science Foundation of China (grant no. 11804287).\\
\indent Baoru Pan and Pan Zhou contributed equally to this work.\\
\indent \emph{Note added.} After completing the GST for altermagnetism in bilayer systems, we learned that related works about altermagnetic bilayers were carried out by Cheng-Cheng Liu's\cite{arXiv1} and Yu-Jun Zhao's\cite{arXiv2} groups.
\bibliography{references}

\begin{thebibliography}{56}%
\makeatletter
\providecommand \@ifxundefined [1]{%
 \@ifx{#1\undefined}
}%
\providecommand \@ifnum [1]{%
 \ifnum #1\expandafter \@firstoftwo
 \else \expandafter \@secondoftwo
 \fi
}%
\providecommand \@ifx [1]{%
 \ifx #1\expandafter \@firstoftwo
 \else \expandafter \@secondoftwo
 \fi
}%
\providecommand \natexlab [1]{#1}%
\providecommand \enquote  [1]{``#1''}%
\providecommand \bibnamefont  [1]{#1}%
\providecommand \bibfnamefont [1]{#1}%
\providecommand \citenamefont [1]{#1}%
\providecommand \href@noop [0]{\@secondoftwo}%
\providecommand \href [0]{\begingroup \@sanitize@url \@href}%
\providecommand \@href[1]{\@@startlink{#1}\@@href}%
\providecommand \@@href[1]{\endgroup#1\@@endlink}%
\providecommand \@sanitize@url [0]{\catcode `\\12\catcode `\$12\catcode
  `\&12\catcode `\#12\catcode `\^12\catcode `\_12\catcode `\%12\relax}%
\providecommand \@@startlink[1]{}%
\providecommand \@@endlink[0]{}%
\providecommand \url  [0]{\begingroup\@sanitize@url \@url }%
\providecommand \@url [1]{\endgroup\@href {#1}{\urlprefix }}%
\providecommand \urlprefix  [0]{URL }%
\providecommand \Eprint [0]{\href }%
\providecommand \doibase [0]{https://doi.org/}%
\providecommand \selectlanguage [0]{\@gobble}%
\providecommand \bibinfo  [0]{\@secondoftwo}%
\providecommand \bibfield  [0]{\@secondoftwo}%
\providecommand \translation [1]{[#1]}%
\providecommand \BibitemOpen [0]{}%
\providecommand \bibitemStop [0]{}%
\providecommand \bibitemNoStop [0]{.\EOS\space}%
\providecommand \EOS [0]{\spacefactor3000\relax}%
\providecommand \BibitemShut  [1]{\csname bibitem#1\endcsname}%
\let\auto@bib@innerbib\@empty
\bibitem [{\citenamefont {{\v{S}}mejkal}\ \emph
  {et~al.}(2022{\natexlab{a}})\citenamefont {{\v{S}}mejkal}, \citenamefont
  {Sinova},\ and\ \citenamefont {Jungwirth}}]{alter1}%
  \BibitemOpen
  \bibfield  {author} {\bibinfo {author} {\bibfnamefont {L.}~\bibnamefont
  {{\v{S}}mejkal}}, \bibinfo {author} {\bibfnamefont {J.}~\bibnamefont
  {Sinova}},\ and\ \bibinfo {author} {\bibfnamefont {T.}~\bibnamefont
  {Jungwirth}},\ }\bibfield  {title} {\bibinfo {title} {Beyond conventional
  ferromagnetism and antiferromagnetism: A phase with nonrelativistic spin and
  crystal rotation symmetry},\ }\href@noop {} {\bibfield  {journal} {\bibinfo
  {journal} {Phys. Rev. X}\ }\textbf {\bibinfo {volume} {12}},\ \bibinfo
  {pages} {031042} (\bibinfo {year} {2022}{\natexlab{a}})}\BibitemShut
  {NoStop}%
\bibitem [{\citenamefont {{\v{S}}mejkal}\ \emph
  {et~al.}(2022{\natexlab{b}})\citenamefont {{\v{S}}mejkal}, \citenamefont
  {Sinova},\ and\ \citenamefont {Jungwirth}}]{alter2}%
  \BibitemOpen
  \bibfield  {author} {\bibinfo {author} {\bibfnamefont {L.}~\bibnamefont
  {{\v{S}}mejkal}}, \bibinfo {author} {\bibfnamefont {J.}~\bibnamefont
  {Sinova}},\ and\ \bibinfo {author} {\bibfnamefont {T.}~\bibnamefont
  {Jungwirth}},\ }\bibfield  {title} {\bibinfo {title} {Emerging research
  landscape of altermagnetism},\ }\href@noop {} {\bibfield  {journal} {\bibinfo
   {journal} {Phys. Rev. X}\ }\textbf {\bibinfo {volume} {12}},\ \bibinfo
  {pages} {040501} (\bibinfo {year} {2022}{\natexlab{b}})}\BibitemShut
  {NoStop}%
\bibitem [{\citenamefont {Mazin}\ \emph {et~al.}(2022)\citenamefont {Mazin}
  \emph {et~al.}}]{alter3}%
  \BibitemOpen
  \bibfield  {author} {\bibinfo {author} {\bibfnamefont {I.}~\bibnamefont
  {Mazin}} \emph {et~al.},\ }\bibfield  {title} {\bibinfo {title}
  {Altermagnetism---a new punch line of fundamental magnetism},\ }\href@noop {}
  {\bibfield  {journal} {\bibinfo  {journal} {Phys. Rev. X}\ }\textbf {\bibinfo
  {volume} {12}},\ \bibinfo {pages} {040002} (\bibinfo {year}
  {2022})}\BibitemShut {NoStop}%
\bibitem [{\citenamefont {Feng}\ \emph {et~al.}(2022)\citenamefont {Feng},
  \citenamefont {Zhou}, \citenamefont {{\v{S}}mejkal}, \citenamefont {Wu},
  \citenamefont {Zhu}, \citenamefont {Guo}, \citenamefont
  {Gonz{\'a}lez-Hern{\'a}ndez}, \citenamefont {Wang}, \citenamefont {Yan},
  \citenamefont {Qin} \emph {et~al.}}]{alter4}%
  \BibitemOpen
  \bibfield  {author} {\bibinfo {author} {\bibfnamefont {Z.}~\bibnamefont
  {Feng}}, \bibinfo {author} {\bibfnamefont {X.}~\bibnamefont {Zhou}}, \bibinfo
  {author} {\bibfnamefont {L.}~\bibnamefont {{\v{S}}mejkal}}, \bibinfo {author}
  {\bibfnamefont {L.}~\bibnamefont {Wu}}, \bibinfo {author} {\bibfnamefont
  {Z.}~\bibnamefont {Zhu}}, \bibinfo {author} {\bibfnamefont {H.}~\bibnamefont
  {Guo}}, \bibinfo {author} {\bibfnamefont {R.}~\bibnamefont
  {Gonz{\'a}lez-Hern{\'a}ndez}}, \bibinfo {author} {\bibfnamefont
  {X.}~\bibnamefont {Wang}}, \bibinfo {author} {\bibfnamefont {H.}~\bibnamefont
  {Yan}}, \bibinfo {author} {\bibfnamefont {P.}~\bibnamefont {Qin}}, \emph
  {et~al.},\ }\bibfield  {title} {\bibinfo {title} {An anomalous hall effect in
  altermagnetic ruthenium dioxide},\ }\href@noop {} {\bibfield  {journal}
  {\bibinfo  {journal} {Nat. Electron.}\ }\textbf {\bibinfo {volume} {5}},\
  \bibinfo {pages} {735} (\bibinfo {year} {2022})}\BibitemShut {NoStop}%
\bibitem [{\citenamefont {Mazin}(2023)}]{alter5}%
  \BibitemOpen
  \bibfield  {author} {\bibinfo {author} {\bibfnamefont {I.}~\bibnamefont
  {Mazin}},\ }\bibfield  {title} {\bibinfo {title} {Altermagnetism in mnte:
  Origin, predicted manifestations, and routes to detwinning},\ }\href@noop {}
  {\bibfield  {journal} {\bibinfo  {journal} {Phys. Rev. B}\ }\textbf {\bibinfo
  {volume} {107}},\ \bibinfo {pages} {L100418} (\bibinfo {year}
  {2023})}\BibitemShut {NoStop}%
\bibitem [{\citenamefont {Maier}\ and\ \citenamefont {Okamoto}(2023)}]{alter6}%
  \BibitemOpen
  \bibfield  {author} {\bibinfo {author} {\bibfnamefont {T.~A.}\ \bibnamefont
  {Maier}}\ and\ \bibinfo {author} {\bibfnamefont {S.}~\bibnamefont
  {Okamoto}},\ }\bibfield  {title} {\bibinfo {title} {Weak-coupling theory of
  neutron scattering as a probe of altermagnetism},\ }\href@noop {} {\bibfield
  {journal} {\bibinfo  {journal} {Phys. Rev. B}\ }\textbf {\bibinfo {volume}
  {108}},\ \bibinfo {pages} {L100402} (\bibinfo {year} {2023})}\BibitemShut
  {NoStop}%
\bibitem [{\citenamefont {Naka}\ \emph {et~al.}(2019)\citenamefont {Naka},
  \citenamefont {Hayami}, \citenamefont {Kusunose}, \citenamefont {Yanagi},
  \citenamefont {Motome},\ and\ \citenamefont {Seo}}]{spintronics1}%
  \BibitemOpen
  \bibfield  {author} {\bibinfo {author} {\bibfnamefont {M.}~\bibnamefont
  {Naka}}, \bibinfo {author} {\bibfnamefont {S.}~\bibnamefont {Hayami}},
  \bibinfo {author} {\bibfnamefont {H.}~\bibnamefont {Kusunose}}, \bibinfo
  {author} {\bibfnamefont {Y.}~\bibnamefont {Yanagi}}, \bibinfo {author}
  {\bibfnamefont {Y.}~\bibnamefont {Motome}},\ and\ \bibinfo {author}
  {\bibfnamefont {H.}~\bibnamefont {Seo}},\ }\bibfield  {title} {\bibinfo
  {title} {Spin current generation in organic antiferromagnets},\ }\href@noop
  {} {\bibfield  {journal} {\bibinfo  {journal} {Nat. Commun.}\ }\textbf
  {\bibinfo {volume} {10}},\ \bibinfo {pages} {4305} (\bibinfo {year}
  {2019})}\BibitemShut {NoStop}%
\bibitem [{\citenamefont {Shao}\ \emph {et~al.}(2021)\citenamefont {Shao},
  \citenamefont {Zhang}, \citenamefont {Li}, \citenamefont {Eom},\ and\
  \citenamefont {Tsymbal}}]{spintronics2}%
  \BibitemOpen
  \bibfield  {author} {\bibinfo {author} {\bibfnamefont {D.-F.}\ \bibnamefont
  {Shao}}, \bibinfo {author} {\bibfnamefont {S.-H.}\ \bibnamefont {Zhang}},
  \bibinfo {author} {\bibfnamefont {M.}~\bibnamefont {Li}}, \bibinfo {author}
  {\bibfnamefont {C.-B.}\ \bibnamefont {Eom}},\ and\ \bibinfo {author}
  {\bibfnamefont {E.~Y.}\ \bibnamefont {Tsymbal}},\ }\bibfield  {title}
  {\bibinfo {title} {Spin-neutral currents for spintronics},\ }\href@noop {}
  {\bibfield  {journal} {\bibinfo  {journal} {Nat. Commun.}\ }\textbf {\bibinfo
  {volume} {12}},\ \bibinfo {pages} {7061} (\bibinfo {year}
  {2021})}\BibitemShut {NoStop}%
\bibitem [{\citenamefont {Ma}\ \emph {et~al.}(2021)\citenamefont {Ma},
  \citenamefont {Hu}, \citenamefont {Li}, \citenamefont {Liu}, \citenamefont
  {Yao}, \citenamefont {Jia},\ and\ \citenamefont {Liu}}]{spintronics3}%
  \BibitemOpen
  \bibfield  {author} {\bibinfo {author} {\bibfnamefont {H.-Y.}\ \bibnamefont
  {Ma}}, \bibinfo {author} {\bibfnamefont {M.}~\bibnamefont {Hu}}, \bibinfo
  {author} {\bibfnamefont {N.}~\bibnamefont {Li}}, \bibinfo {author}
  {\bibfnamefont {J.}~\bibnamefont {Liu}}, \bibinfo {author} {\bibfnamefont
  {W.}~\bibnamefont {Yao}}, \bibinfo {author} {\bibfnamefont {J.-F.}\
  \bibnamefont {Jia}},\ and\ \bibinfo {author} {\bibfnamefont {J.}~\bibnamefont
  {Liu}},\ }\bibfield  {title} {\bibinfo {title} {Multifunctional
  antiferromagnetic materials with giant piezomagnetism and noncollinear spin
  current},\ }\href@noop {} {\bibfield  {journal} {\bibinfo  {journal} {Nat.
  Commun.}\ }\textbf {\bibinfo {volume} {12}},\ \bibinfo {pages} {2846}
  (\bibinfo {year} {2021})}\BibitemShut {NoStop}%
\bibitem [{\citenamefont {Naka}\ \emph {et~al.}(2021)\citenamefont {Naka},
  \citenamefont {Motome},\ and\ \citenamefont {Seo}}]{spintronics4}%
  \BibitemOpen
  \bibfield  {author} {\bibinfo {author} {\bibfnamefont {M.}~\bibnamefont
  {Naka}}, \bibinfo {author} {\bibfnamefont {Y.}~\bibnamefont {Motome}},\ and\
  \bibinfo {author} {\bibfnamefont {H.}~\bibnamefont {Seo}},\ }\bibfield
  {title} {\bibinfo {title} {Perovskite as a spin current generator},\
  }\href@noop {} {\bibfield  {journal} {\bibinfo  {journal} {Phys. Rev. B}\
  }\textbf {\bibinfo {volume} {103}},\ \bibinfo {pages} {125114} (\bibinfo
  {year} {2021})}\BibitemShut {NoStop}%
\bibitem [{\citenamefont {{\v{S}}mejkal}\ \emph
  {et~al.}(2022{\natexlab{c}})\citenamefont {{\v{S}}mejkal}, \citenamefont
  {Hellenes}, \citenamefont {Gonz{\'a}lez-Hern{\'a}ndez}, \citenamefont
  {Sinova},\ and\ \citenamefont {Jungwirth}}]{spintronics5}%
  \BibitemOpen
  \bibfield  {author} {\bibinfo {author} {\bibfnamefont {L.}~\bibnamefont
  {{\v{S}}mejkal}}, \bibinfo {author} {\bibfnamefont {A.~B.}\ \bibnamefont
  {Hellenes}}, \bibinfo {author} {\bibfnamefont {R.}~\bibnamefont
  {Gonz{\'a}lez-Hern{\'a}ndez}}, \bibinfo {author} {\bibfnamefont
  {J.}~\bibnamefont {Sinova}},\ and\ \bibinfo {author} {\bibfnamefont
  {T.}~\bibnamefont {Jungwirth}},\ }\bibfield  {title} {\bibinfo {title} {Giant
  and tunneling magnetoresistance in unconventional collinear antiferromagnets
  with nonrelativistic spin-momentum coupling},\ }\href@noop {} {\bibfield
  {journal} {\bibinfo  {journal} {Phys. Rev. X}\ }\textbf {\bibinfo {volume}
  {12}},\ \bibinfo {pages} {011028} (\bibinfo {year}
  {2022}{\natexlab{c}})}\BibitemShut {NoStop}%
\bibitem [{\citenamefont {Bauer}\ \emph {et~al.}(2012)\citenamefont {Bauer},
  \citenamefont {Saitoh},\ and\ \citenamefont {Van~Wees}}]{caloritronics}%
  \BibitemOpen
  \bibfield  {author} {\bibinfo {author} {\bibfnamefont {G.~E.}\ \bibnamefont
  {Bauer}}, \bibinfo {author} {\bibfnamefont {E.}~\bibnamefont {Saitoh}},\ and\
  \bibinfo {author} {\bibfnamefont {B.~J.}\ \bibnamefont {Van~Wees}},\
  }\bibfield  {title} {\bibinfo {title} {Spin caloritronics},\ }\href@noop {}
  {\bibfield  {journal} {\bibinfo  {journal} {Nat. Mater.}\ }\textbf {\bibinfo
  {volume} {11}},\ \bibinfo {pages} {391} (\bibinfo {year} {2012})}\BibitemShut
  {NoStop}%
\bibitem [{\citenamefont {Ramesh}\ and\ \citenamefont
  {Spaldin}(2007)}]{multiferroics1}%
  \BibitemOpen
  \bibfield  {author} {\bibinfo {author} {\bibfnamefont {R.}~\bibnamefont
  {Ramesh}}\ and\ \bibinfo {author} {\bibfnamefont {N.~A.}\ \bibnamefont
  {Spaldin}},\ }\bibfield  {title} {\bibinfo {title} {Multiferroics: progress
  and prospects in thin films},\ }\href@noop {} {\bibfield  {journal} {\bibinfo
   {journal} {Nat. Mater.}\ }\textbf {\bibinfo {volume} {6}},\ \bibinfo {pages}
  {21} (\bibinfo {year} {2007})}\BibitemShut {NoStop}%
\bibitem [{\citenamefont {Bhattacharjee}\ \emph {et~al.}(2009)\citenamefont
  {Bhattacharjee}, \citenamefont {Bousquet},\ and\ \citenamefont
  {Ghosez}}]{multiferroics2}%
  \BibitemOpen
  \bibfield  {author} {\bibinfo {author} {\bibfnamefont {S.}~\bibnamefont
  {Bhattacharjee}}, \bibinfo {author} {\bibfnamefont {E.}~\bibnamefont
  {Bousquet}},\ and\ \bibinfo {author} {\bibfnamefont {P.}~\bibnamefont
  {Ghosez}},\ }\bibfield  {title} {\bibinfo {title} {Engineering multiferroism
  in {CaMnO$_{3}$}},\ }\href@noop {} {\bibfield  {journal} {\bibinfo  {journal}
  {Phys. Rev. Lett.}\ }\textbf {\bibinfo {volume} {102}},\ \bibinfo {pages}
  {117602} (\bibinfo {year} {2009})}\BibitemShut {NoStop}%
\bibitem [{\citenamefont {Sigrist}\ and\ \citenamefont
  {Ueda}(1991)}]{superconductivity1}%
  \BibitemOpen
  \bibfield  {author} {\bibinfo {author} {\bibfnamefont {M.}~\bibnamefont
  {Sigrist}}\ and\ \bibinfo {author} {\bibfnamefont {K.}~\bibnamefont {Ueda}},\
  }\bibfield  {title} {\bibinfo {title} {Phenomenological theory of
  unconventional superconductivity},\ }\href@noop {} {\bibfield  {journal}
  {\bibinfo  {journal} {Rev. Mod. Phys.}\ }\textbf {\bibinfo {volume} {63}},\
  \bibinfo {pages} {239} (\bibinfo {year} {1991})}\BibitemShut {NoStop}%
\bibitem [{\citenamefont {Mazin}\ \emph {et~al.}(2001)\citenamefont {Mazin},
  \citenamefont {Golubov},\ and\ \citenamefont
  {Nadgorny}}]{superconductivity2}%
  \BibitemOpen
  \bibfield  {author} {\bibinfo {author} {\bibfnamefont {I.}~\bibnamefont
  {Mazin}}, \bibinfo {author} {\bibfnamefont {A.~A.}\ \bibnamefont {Golubov}},\
  and\ \bibinfo {author} {\bibfnamefont {B.}~\bibnamefont {Nadgorny}},\
  }\bibfield  {title} {\bibinfo {title} {Probing spin polarization with andreev
  reflection: A theoretical basis},\ }\href@noop {} {\bibfield  {journal}
  {\bibinfo  {journal} {J. Appl. Phys.}\ }\textbf {\bibinfo {volume} {89}},\
  \bibinfo {pages} {7576} (\bibinfo {year} {2001})}\BibitemShut {NoStop}%
\bibitem [{\citenamefont {Flensberg}\ \emph {et~al.}(2021)\citenamefont
  {Flensberg}, \citenamefont {von Oppen},\ and\ \citenamefont
  {Stern}}]{superconductivity3}%
  \BibitemOpen
  \bibfield  {author} {\bibinfo {author} {\bibfnamefont {K.}~\bibnamefont
  {Flensberg}}, \bibinfo {author} {\bibfnamefont {F.}~\bibnamefont {von
  Oppen}},\ and\ \bibinfo {author} {\bibfnamefont {A.}~\bibnamefont {Stern}},\
  }\bibfield  {title} {\bibinfo {title} {Engineered platforms for topological
  superconductivity and majorana zero modes},\ }\href@noop {} {\bibfield
  {journal} {\bibinfo  {journal} {Nat. Rev. Mater.}\ }\textbf {\bibinfo
  {volume} {6}},\ \bibinfo {pages} {944} (\bibinfo {year} {2021})}\BibitemShut
  {NoStop}%
\bibitem [{\citenamefont {Mazin}(2022)}]{superconductivity4}%
  \BibitemOpen
  \bibfield  {author} {\bibinfo {author} {\bibfnamefont {I.~I.}\ \bibnamefont
  {Mazin}},\ }\bibfield  {title} {\bibinfo {title} {Notes on altermagnetism and
  superconductivity},\ }\href@noop {} {\bibfield  {journal} {\bibinfo
  {journal} {arXiv preprint arXiv:2203.05000}\ } (\bibinfo {year}
  {2022})}\BibitemShut {NoStop}%
\bibitem [{\citenamefont {{\v{S}}mejkal}\ \emph {et~al.}(2020)\citenamefont
  {{\v{S}}mejkal}, \citenamefont {Gonz{\'a}lez-Hern{\'a}ndez}, \citenamefont
  {Jungwirth},\ and\ \citenamefont {Sinova}}]{RuO2_1}%
  \BibitemOpen
  \bibfield  {author} {\bibinfo {author} {\bibfnamefont {L.}~\bibnamefont
  {{\v{S}}mejkal}}, \bibinfo {author} {\bibfnamefont {R.}~\bibnamefont
  {Gonz{\'a}lez-Hern{\'a}ndez}}, \bibinfo {author} {\bibfnamefont
  {T.}~\bibnamefont {Jungwirth}},\ and\ \bibinfo {author} {\bibfnamefont
  {J.}~\bibnamefont {Sinova}},\ }\bibfield  {title} {\bibinfo {title} {Crystal
  time-reversal symmetry breaking and spontaneous hall effect in collinear
  antiferromagnets},\ }\href@noop {} {\bibfield  {journal} {\bibinfo  {journal}
  {Sci. Adv.}\ }\textbf {\bibinfo {volume} {6}},\ \bibinfo {pages} {eaaz8809}
  (\bibinfo {year} {2020})}\BibitemShut {NoStop}%
\bibitem [{\citenamefont {Gonz{\'a}lez-Hern{\'a}ndez}\ \emph
  {et~al.}(2021)\citenamefont {Gonz{\'a}lez-Hern{\'a}ndez}, \citenamefont
  {{\v{S}}mejkal}, \citenamefont {V{\`y}born{\`y}}, \citenamefont {Yahagi},
  \citenamefont {Sinova}, \citenamefont {Jungwirth},\ and\ \citenamefont
  {{\v{Z}}elezn{\`y}}}]{RuO2_2}%
  \BibitemOpen
  \bibfield  {author} {\bibinfo {author} {\bibfnamefont {R.}~\bibnamefont
  {Gonz{\'a}lez-Hern{\'a}ndez}}, \bibinfo {author} {\bibfnamefont
  {L.}~\bibnamefont {{\v{S}}mejkal}}, \bibinfo {author} {\bibfnamefont
  {K.}~\bibnamefont {V{\`y}born{\`y}}}, \bibinfo {author} {\bibfnamefont
  {Y.}~\bibnamefont {Yahagi}}, \bibinfo {author} {\bibfnamefont
  {J.}~\bibnamefont {Sinova}}, \bibinfo {author} {\bibfnamefont
  {T.}~\bibnamefont {Jungwirth}},\ and\ \bibinfo {author} {\bibfnamefont
  {J.}~\bibnamefont {{\v{Z}}elezn{\`y}}},\ }\bibfield  {title} {\bibinfo
  {title} {Efficient electrical spin splitter based on nonrelativistic
  collinear antiferromagnetism},\ }\href@noop {} {\bibfield  {journal}
  {\bibinfo  {journal} {Phys. Rev. Lett.}\ }\textbf {\bibinfo {volume} {126}},\
  \bibinfo {pages} {127701} (\bibinfo {year} {2021})}\BibitemShut {NoStop}%
\bibitem [{\citenamefont {Ahn}\ \emph {et~al.}(2019)\citenamefont {Ahn},
  \citenamefont {Hariki}, \citenamefont {Lee},\ and\ \citenamefont
  {Kune{\v{s}}}}]{RuO2_3}%
  \BibitemOpen
  \bibfield  {author} {\bibinfo {author} {\bibfnamefont {K.-H.}\ \bibnamefont
  {Ahn}}, \bibinfo {author} {\bibfnamefont {A.}~\bibnamefont {Hariki}},
  \bibinfo {author} {\bibfnamefont {K.-W.}\ \bibnamefont {Lee}},\ and\ \bibinfo
  {author} {\bibfnamefont {J.}~\bibnamefont {Kune{\v{s}}}},\ }\bibfield
  {title} {\bibinfo {title} {Antiferromagnetism in {RuO$_{2}$} as d-wave
  pomeranchuk instability},\ }\href@noop {} {\bibfield  {journal} {\bibinfo
  {journal} {Phys. Rev. B}\ }\textbf {\bibinfo {volume} {99}},\ \bibinfo
  {pages} {184432} (\bibinfo {year} {2019})}\BibitemShut {NoStop}%
\bibitem [{\citenamefont {Bose}\ \emph {et~al.}(2022)\citenamefont {Bose},
  \citenamefont {Schreiber}, \citenamefont {Jain}, \citenamefont {Shao},
  \citenamefont {Nair}, \citenamefont {Sun}, \citenamefont {Zhang},
  \citenamefont {Muller}, \citenamefont {Tsymbal}, \citenamefont {Schlom} \emph
  {et~al.}}]{spintronics6}%
  \BibitemOpen
  \bibfield  {author} {\bibinfo {author} {\bibfnamefont {A.}~\bibnamefont
  {Bose}}, \bibinfo {author} {\bibfnamefont {N.~J.}\ \bibnamefont {Schreiber}},
  \bibinfo {author} {\bibfnamefont {R.}~\bibnamefont {Jain}}, \bibinfo {author}
  {\bibfnamefont {D.-F.}\ \bibnamefont {Shao}}, \bibinfo {author}
  {\bibfnamefont {H.~P.}\ \bibnamefont {Nair}}, \bibinfo {author}
  {\bibfnamefont {J.}~\bibnamefont {Sun}}, \bibinfo {author} {\bibfnamefont
  {X.~S.}\ \bibnamefont {Zhang}}, \bibinfo {author} {\bibfnamefont {D.~A.}\
  \bibnamefont {Muller}}, \bibinfo {author} {\bibfnamefont {E.~Y.}\
  \bibnamefont {Tsymbal}}, \bibinfo {author} {\bibfnamefont {D.~G.}\
  \bibnamefont {Schlom}}, \emph {et~al.},\ }\bibfield  {title} {\bibinfo
  {title} {Tilted spin current generated by the collinear antiferromagnet
  ruthenium dioxide},\ }\href@noop {} {\bibfield  {journal} {\bibinfo
  {journal} {Nat. Electron.}\ }\textbf {\bibinfo {volume} {5}},\ \bibinfo
  {pages} {267} (\bibinfo {year} {2022})}\BibitemShut {NoStop}%
\bibitem [{\citenamefont {Bai}\ \emph {et~al.}(2022)\citenamefont {Bai},
  \citenamefont {Han}, \citenamefont {Feng}, \citenamefont {Zhou},
  \citenamefont {Su}, \citenamefont {Wang}, \citenamefont {Liao}, \citenamefont
  {Zhu}, \citenamefont {Chen}, \citenamefont {Pan} \emph
  {et~al.}}]{spintronics7}%
  \BibitemOpen
  \bibfield  {author} {\bibinfo {author} {\bibfnamefont {H.}~\bibnamefont
  {Bai}}, \bibinfo {author} {\bibfnamefont {L.}~\bibnamefont {Han}}, \bibinfo
  {author} {\bibfnamefont {X.}~\bibnamefont {Feng}}, \bibinfo {author}
  {\bibfnamefont {Y.}~\bibnamefont {Zhou}}, \bibinfo {author} {\bibfnamefont
  {R.}~\bibnamefont {Su}}, \bibinfo {author} {\bibfnamefont {Q.}~\bibnamefont
  {Wang}}, \bibinfo {author} {\bibfnamefont {L.}~\bibnamefont {Liao}}, \bibinfo
  {author} {\bibfnamefont {W.}~\bibnamefont {Zhu}}, \bibinfo {author}
  {\bibfnamefont {X.}~\bibnamefont {Chen}}, \bibinfo {author} {\bibfnamefont
  {F.}~\bibnamefont {Pan}}, \emph {et~al.},\ }\bibfield  {title} {\bibinfo
  {title} {Observation of spin splitting torque in a collinear antiferromagnet
  {RuO$_{2}$}},\ }\href@noop {} {\bibfield  {journal} {\bibinfo  {journal}
  {Phys. Rev. Lett.}\ }\textbf {\bibinfo {volume} {128}},\ \bibinfo {pages}
  {197202} (\bibinfo {year} {2022})}\BibitemShut {NoStop}%
\bibitem [{\citenamefont {He}\ \emph {et~al.}(2023)\citenamefont {He},
  \citenamefont {Wang}, \citenamefont {Luo}, \citenamefont {Zeng},
  \citenamefont {Chen},\ and\ \citenamefont {Tang}}]{tang_liming}%
  \BibitemOpen
  \bibfield  {author} {\bibinfo {author} {\bibfnamefont {R.}~\bibnamefont
  {He}}, \bibinfo {author} {\bibfnamefont {D.}~\bibnamefont {Wang}}, \bibinfo
  {author} {\bibfnamefont {N.}~\bibnamefont {Luo}}, \bibinfo {author}
  {\bibfnamefont {J.}~\bibnamefont {Zeng}}, \bibinfo {author} {\bibfnamefont
  {K.-Q.}\ \bibnamefont {Chen}},\ and\ \bibinfo {author} {\bibfnamefont
  {L.-M.}\ \bibnamefont {Tang}},\ }\bibfield  {title} {\bibinfo {title}
  {Nonrelativistic spin-momentum coupling in antiferromagnetic twisted
  bilayers},\ }\href@noop {} {\bibfield  {journal} {\bibinfo  {journal} {Phys.
  Rev. Lett.}\ }\textbf {\bibinfo {volume} {130}},\ \bibinfo {pages} {046401}
  (\bibinfo {year} {2023})}\BibitemShut {NoStop}%
\bibitem [{sup()}]{sup}%
  \BibitemOpen
  \href@noop {} {\bibinfo {title} {\textit{See Supplemental Material for matrix
  representations of the symmetry operations with bilayer coordinate system,
  single layer point groups, stacking operator, bilayer point groups, and
  whether it can form altermagnetic bilayers, detailed discussions about
  twisted bilayer CrBrSe and NiZrI$_{6}$, energy difference between interlayer
  antiferromagnetic and other magnetic states, information of twisted CrBrSe
  with different Stacking angles, effect of interlayer interaction, effect of
  translation to bilayer NiZrI$_{6}$, other representative candidate materials,
  the choice of the origin, discussion about the coexist of altermagnetism and
  piezoelectricity, which includes Refs. \cite{tang_liming, xiang_hj, alter1,
  CrSBr_1, CrSBr_2, twist_control1, twist_control2, PES, Stacking-exp1,
  Stacking-exp2, Stacking-exp3, Stacking-exp4, pie_BN, pie_crs2, pie_graphene,
  pie_tmdc, pie_2d, imaginary1, imaginary2, imaginary3}.}}}\BibitemShut {Stop}%
\bibitem [{\citenamefont {Brinkman}\ and\ \citenamefont
  {Elliott}(1966)}]{spin_space_1}%
  \BibitemOpen
  \bibfield  {author} {\bibinfo {author} {\bibfnamefont {W.}~\bibnamefont
  {Brinkman}}\ and\ \bibinfo {author} {\bibfnamefont {R.~J.}\ \bibnamefont
  {Elliott}},\ }\bibfield  {title} {\bibinfo {title} {Theory of spin-space
  groups},\ }\href@noop {} {\bibfield  {journal} {\bibinfo  {journal} {Proc. R.
  Soc. A}\ }\textbf {\bibinfo {volume} {294}},\ \bibinfo {pages} {343}
  (\bibinfo {year} {1966})}\BibitemShut {NoStop}%
\bibitem [{\citenamefont {Litvin}\ and\ \citenamefont
  {Opechowski}(1974)}]{spin_space_2}%
  \BibitemOpen
  \bibfield  {author} {\bibinfo {author} {\bibfnamefont {D.~B.}\ \bibnamefont
  {Litvin}}\ and\ \bibinfo {author} {\bibfnamefont {W.}~\bibnamefont
  {Opechowski}},\ }\bibfield  {title} {\bibinfo {title} {Spin groups},\
  }\href@noop {} {\bibfield  {journal} {\bibinfo  {journal} {Physica}\ }\textbf
  {\bibinfo {volume} {76}},\ \bibinfo {pages} {538} (\bibinfo {year}
  {1974})}\BibitemShut {NoStop}%
\bibitem [{\citenamefont {Litvin}(1977)}]{spin_space_3}%
  \BibitemOpen
  \bibfield  {author} {\bibinfo {author} {\bibfnamefont {D.~B.}\ \bibnamefont
  {Litvin}},\ }\bibfield  {title} {\bibinfo {title} {Spin point groups},\
  }\href@noop {} {\bibfield  {journal} {\bibinfo  {journal} {Acta Crystallogr.,
  Sect. A}\ }\textbf {\bibinfo {volume} {33}},\ \bibinfo {pages} {279}
  (\bibinfo {year} {1977})}\BibitemShut {NoStop}%
\bibitem [{\citenamefont {Yuan}\ \emph {et~al.}(2023)\citenamefont {Yuan},
  \citenamefont {Zhang}, \citenamefont {Acosta},\ and\ \citenamefont
  {Zunger}}]{hidden_spin}%
  \BibitemOpen
  \bibfield  {author} {\bibinfo {author} {\bibfnamefont {L.-D.}\ \bibnamefont
  {Yuan}}, \bibinfo {author} {\bibfnamefont {X.}~\bibnamefont {Zhang}},
  \bibinfo {author} {\bibfnamefont {C.~M.}\ \bibnamefont {Acosta}},\ and\
  \bibinfo {author} {\bibfnamefont {A.}~\bibnamefont {Zunger}},\ }\bibfield
  {title} {\bibinfo {title} {Uncovering spin-orbit coupling-independent hidden
  spin polarization of energy bands in antiferromagnets},\ }\href@noop {}
  {\bibfield  {journal} {\bibinfo  {journal} {Nat. Commun.}\ }\textbf {\bibinfo
  {volume} {14}},\ \bibinfo {pages} {5301} (\bibinfo {year}
  {2023})}\BibitemShut {NoStop}%
\bibitem [{\citenamefont {Gratias}\ and\ \citenamefont
  {Quiquandon}(2023)}]{twisted_bilayers}%
  \BibitemOpen
  \bibfield  {author} {\bibinfo {author} {\bibfnamefont {D.}~\bibnamefont
  {Gratias}}\ and\ \bibinfo {author} {\bibfnamefont {M.}~\bibnamefont
  {Quiquandon}},\ }\bibfield  {title} {\bibinfo {title} {Crystallography of
  homophase twisted bilayers: coincidence, union lattices and space groups},\
  }\href@noop {} {\bibfield  {journal} {\bibinfo  {journal} {Acta Cryst. A}\
  }\textbf {\bibinfo {volume} {79}} (\bibinfo {year} {2023})}\BibitemShut
  {NoStop}%
\bibitem [{\citenamefont {Gjerding}\ \emph {et~al.}(2021)\citenamefont
  {Gjerding}, \citenamefont {Taghizadeh}, \citenamefont {Rasmussen},
  \citenamefont {Ali}, \citenamefont {Bertoldo}, \citenamefont {Deilmann},
  \citenamefont {Kn{\o}sgaard}, \citenamefont {Kruse}, \citenamefont {Larsen},
  \citenamefont {Manti} \emph {et~al.}}]{c2db}%
  \BibitemOpen
  \bibfield  {author} {\bibinfo {author} {\bibfnamefont {M.~N.}\ \bibnamefont
  {Gjerding}}, \bibinfo {author} {\bibfnamefont {A.}~\bibnamefont
  {Taghizadeh}}, \bibinfo {author} {\bibfnamefont {A.}~\bibnamefont
  {Rasmussen}}, \bibinfo {author} {\bibfnamefont {S.}~\bibnamefont {Ali}},
  \bibinfo {author} {\bibfnamefont {F.}~\bibnamefont {Bertoldo}}, \bibinfo
  {author} {\bibfnamefont {T.}~\bibnamefont {Deilmann}}, \bibinfo {author}
  {\bibfnamefont {N.~R.}\ \bibnamefont {Kn{\o}sgaard}}, \bibinfo {author}
  {\bibfnamefont {M.}~\bibnamefont {Kruse}}, \bibinfo {author} {\bibfnamefont
  {A.~H.}\ \bibnamefont {Larsen}}, \bibinfo {author} {\bibfnamefont
  {S.}~\bibnamefont {Manti}}, \emph {et~al.},\ }\bibfield  {title} {\bibinfo
  {title} {Recent progress of the computational {2D} materials database
  {(C2DB)}},\ }\href@noop {} {\bibfield  {journal} {\bibinfo  {journal} {2D
  Materials}\ }\textbf {\bibinfo {volume} {8}},\ \bibinfo {pages} {044002}
  (\bibinfo {year} {2021})}\BibitemShut {NoStop}%
\bibitem [{\citenamefont {Jiang}\ \emph {et~al.}(2019)\citenamefont {Jiang},
  \citenamefont {Wang}, \citenamefont {Chen}, \citenamefont {Zhong},
  \citenamefont {Yuan}, \citenamefont {Lu},\ and\ \citenamefont
  {Ji}}]{tunable_mag1}%
  \BibitemOpen
  \bibfield  {author} {\bibinfo {author} {\bibfnamefont {P.}~\bibnamefont
  {Jiang}}, \bibinfo {author} {\bibfnamefont {C.}~\bibnamefont {Wang}},
  \bibinfo {author} {\bibfnamefont {D.}~\bibnamefont {Chen}}, \bibinfo {author}
  {\bibfnamefont {Z.}~\bibnamefont {Zhong}}, \bibinfo {author} {\bibfnamefont
  {Z.}~\bibnamefont {Yuan}}, \bibinfo {author} {\bibfnamefont {Z.-Y.}\
  \bibnamefont {Lu}},\ and\ \bibinfo {author} {\bibfnamefont {W.}~\bibnamefont
  {Ji}},\ }\bibfield  {title} {\bibinfo {title} {Stacking tunable interlayer
  magnetism in bilayer {CrI$_3$}},\ }\href@noop {} {\bibfield  {journal}
  {\bibinfo  {journal} {Phys. Rev. B}\ }\textbf {\bibinfo {volume} {99}},\
  \bibinfo {pages} {144401} (\bibinfo {year} {2019})}\BibitemShut {NoStop}%
\bibitem [{\citenamefont {Sivadas}\ \emph {et~al.}(2018)\citenamefont
  {Sivadas}, \citenamefont {Okamoto}, \citenamefont {Xu}, \citenamefont
  {Fennie},\ and\ \citenamefont {Xiao}}]{tunable_mag2}%
  \BibitemOpen
  \bibfield  {author} {\bibinfo {author} {\bibfnamefont {N.}~\bibnamefont
  {Sivadas}}, \bibinfo {author} {\bibfnamefont {S.}~\bibnamefont {Okamoto}},
  \bibinfo {author} {\bibfnamefont {X.}~\bibnamefont {Xu}}, \bibinfo {author}
  {\bibfnamefont {C.~J.}\ \bibnamefont {Fennie}},\ and\ \bibinfo {author}
  {\bibfnamefont {D.}~\bibnamefont {Xiao}},\ }\bibfield  {title} {\bibinfo
  {title} {Stacking-dependent magnetism in bilayer {CrI$_{3}$}},\ }\href@noop
  {} {\bibfield  {journal} {\bibinfo  {journal} {Nano Lett.}\ }\textbf
  {\bibinfo {volume} {18}},\ \bibinfo {pages} {7658} (\bibinfo {year}
  {2018})}\BibitemShut {NoStop}%
\bibitem [{\citenamefont {Jiang}\ \emph
  {et~al.}(2018{\natexlab{a}})\citenamefont {Jiang}, \citenamefont {Li},
  \citenamefont {Wang}, \citenamefont {Mak},\ and\ \citenamefont
  {Shan}}]{tunable_mag4}%
  \BibitemOpen
  \bibfield  {author} {\bibinfo {author} {\bibfnamefont {S.}~\bibnamefont
  {Jiang}}, \bibinfo {author} {\bibfnamefont {L.}~\bibnamefont {Li}}, \bibinfo
  {author} {\bibfnamefont {Z.}~\bibnamefont {Wang}}, \bibinfo {author}
  {\bibfnamefont {K.~F.}\ \bibnamefont {Mak}},\ and\ \bibinfo {author}
  {\bibfnamefont {J.}~\bibnamefont {Shan}},\ }\bibfield  {title} {\bibinfo
  {title} {Controlling magnetism in {2D} {CrI$_{3}$} by electrostatic doping},\
  }\href@noop {} {\bibfield  {journal} {\bibinfo  {journal} {Nat.
  Nanotechnol.}\ }\textbf {\bibinfo {volume} {13}},\ \bibinfo {pages} {549}
  (\bibinfo {year} {2018}{\natexlab{a}})}\BibitemShut {NoStop}%
\bibitem [{\citenamefont {Jiang}\ \emph
  {et~al.}(2018{\natexlab{b}})\citenamefont {Jiang}, \citenamefont {Shan},\
  and\ \citenamefont {Mak}}]{tunable_mag5}%
  \BibitemOpen
  \bibfield  {author} {\bibinfo {author} {\bibfnamefont {S.}~\bibnamefont
  {Jiang}}, \bibinfo {author} {\bibfnamefont {J.}~\bibnamefont {Shan}},\ and\
  \bibinfo {author} {\bibfnamefont {K.~F.}\ \bibnamefont {Mak}},\ }\bibfield
  {title} {\bibinfo {title} {Electric-field switching of two-dimensional van
  der waals magnets},\ }\href@noop {} {\bibfield  {journal} {\bibinfo
  {journal} {Nat. Mater.}\ }\textbf {\bibinfo {volume} {17}},\ \bibinfo {pages}
  {406} (\bibinfo {year} {2018}{\natexlab{b}})}\BibitemShut {NoStop}%
\bibitem [{\citenamefont {Huang}\ \emph {et~al.}(2018)\citenamefont {Huang},
  \citenamefont {Clark}, \citenamefont {Klein}, \citenamefont {MacNeill},
  \citenamefont {Navarro-Moratalla}, \citenamefont {Seyler}, \citenamefont
  {Wilson}, \citenamefont {McGuire}, \citenamefont {Cobden}, \citenamefont
  {Xiao} \emph {et~al.}}]{tunable_mag6}%
  \BibitemOpen
  \bibfield  {author} {\bibinfo {author} {\bibfnamefont {B.}~\bibnamefont
  {Huang}}, \bibinfo {author} {\bibfnamefont {G.}~\bibnamefont {Clark}},
  \bibinfo {author} {\bibfnamefont {D.~R.}\ \bibnamefont {Klein}}, \bibinfo
  {author} {\bibfnamefont {D.}~\bibnamefont {MacNeill}}, \bibinfo {author}
  {\bibfnamefont {E.}~\bibnamefont {Navarro-Moratalla}}, \bibinfo {author}
  {\bibfnamefont {K.~L.}\ \bibnamefont {Seyler}}, \bibinfo {author}
  {\bibfnamefont {N.}~\bibnamefont {Wilson}}, \bibinfo {author} {\bibfnamefont
  {M.~A.}\ \bibnamefont {McGuire}}, \bibinfo {author} {\bibfnamefont {D.~H.}\
  \bibnamefont {Cobden}}, \bibinfo {author} {\bibfnamefont {D.}~\bibnamefont
  {Xiao}}, \emph {et~al.},\ }\bibfield  {title} {\bibinfo {title} {Electrical
  control of {2D} magnetism in bilayer {CrI$_{3}$}},\ }\href@noop {} {\bibfield
   {journal} {\bibinfo  {journal} {Nat. Nanotechnol.}\ }\textbf {\bibinfo
  {volume} {13}},\ \bibinfo {pages} {544} (\bibinfo {year} {2018})}\BibitemShut
  {NoStop}%
\bibitem [{\citenamefont {Liu}\ \emph {et~al.}(2024)\citenamefont {Liu},
  \citenamefont {Yu},\ and\ \citenamefont {Liu}}]{arXiv1}%
  \BibitemOpen
  \bibfield  {author} {\bibinfo {author} {\bibfnamefont {Y.}~\bibnamefont
  {Liu}}, \bibinfo {author} {\bibfnamefont {J.}~\bibnamefont {Yu}},\ and\
  \bibinfo {author} {\bibfnamefont {C.-C.}\ \bibnamefont {Liu}},\ }\bibfield
  {title} {\bibinfo {title} {Twisted magnetic van der waals bilayers: An ideal
  platform for altermagnetism},\ }\href@noop {} {\bibfield  {journal} {\bibinfo
   {journal} {arXiv preprint arXiv:2404.17146}\ } (\bibinfo {year}
  {2024})}\BibitemShut {NoStop}%
\bibitem [{\citenamefont {Zeng}\ and\ \citenamefont {Zhao}(2024)}]{arXiv2}%
  \BibitemOpen
  \bibfield  {author} {\bibinfo {author} {\bibfnamefont {S.}~\bibnamefont
  {Zeng}}\ and\ \bibinfo {author} {\bibfnamefont {Y.-J.}\ \bibnamefont
  {Zhao}},\ }\bibfield  {title} {\bibinfo {title} {Bilayer stacking a-type
  altermagnet: A general approach to generating two-dimensional
  altermagnetism},\ }\href@noop {} {\bibfield  {journal} {\bibinfo  {journal}
  {arXiv preprint arXiv:2407.15097}\ } (\bibinfo {year} {2024})}\BibitemShut
  {NoStop}%
\bibitem [{\citenamefont {Ji}\ \emph {et~al.}(2023)\citenamefont {Ji},
  \citenamefont {Yu}, \citenamefont {Xu},\ and\ \citenamefont
  {Xiang}}]{xiang_hj}%
  \BibitemOpen
  \bibfield  {author} {\bibinfo {author} {\bibfnamefont {J.}~\bibnamefont
  {Ji}}, \bibinfo {author} {\bibfnamefont {G.}~\bibnamefont {Yu}}, \bibinfo
  {author} {\bibfnamefont {C.}~\bibnamefont {Xu}},\ and\ \bibinfo {author}
  {\bibfnamefont {H.}~\bibnamefont {Xiang}},\ }\bibfield  {title} {\bibinfo
  {title} {General theory for bilayer stacking ferroelectricity},\ }\href@noop
  {} {\bibfield  {journal} {\bibinfo  {journal} {Phys. Rev. Lett.}\ }\textbf
  {\bibinfo {volume} {130}},\ \bibinfo {pages} {146801} (\bibinfo {year}
  {2023})}\BibitemShut {NoStop}%
\bibitem [{\citenamefont {Telford}\ \emph {et~al.}(2020)\citenamefont
  {Telford}, \citenamefont {Dismukes}, \citenamefont {Lee}, \citenamefont
  {Cheng}, \citenamefont {Wieteska}, \citenamefont {Bartholomew}, \citenamefont
  {Chen}, \citenamefont {Xu}, \citenamefont {Pasupathy}, \citenamefont {Zhu}
  \emph {et~al.}}]{CrSBr_1}%
  \BibitemOpen
  \bibfield  {author} {\bibinfo {author} {\bibfnamefont {E.~J.}\ \bibnamefont
  {Telford}}, \bibinfo {author} {\bibfnamefont {A.~H.}\ \bibnamefont
  {Dismukes}}, \bibinfo {author} {\bibfnamefont {K.}~\bibnamefont {Lee}},
  \bibinfo {author} {\bibfnamefont {M.}~\bibnamefont {Cheng}}, \bibinfo
  {author} {\bibfnamefont {A.}~\bibnamefont {Wieteska}}, \bibinfo {author}
  {\bibfnamefont {A.~K.}\ \bibnamefont {Bartholomew}}, \bibinfo {author}
  {\bibfnamefont {Y.-S.}\ \bibnamefont {Chen}}, \bibinfo {author}
  {\bibfnamefont {X.}~\bibnamefont {Xu}}, \bibinfo {author} {\bibfnamefont
  {A.~N.}\ \bibnamefont {Pasupathy}}, \bibinfo {author} {\bibfnamefont
  {X.}~\bibnamefont {Zhu}}, \emph {et~al.},\ }\bibfield  {title} {\bibinfo
  {title} {Layered antiferromagnetism induces large negative magnetoresistance
  in the van der waals semiconductor crsbr},\ }\href@noop {} {\bibfield
  {journal} {\bibinfo  {journal} {Adv. Mater.}\ }\textbf {\bibinfo {volume}
  {32}},\ \bibinfo {pages} {2003240} (\bibinfo {year} {2020})}\BibitemShut
  {NoStop}%
\bibitem [{\citenamefont {Klein}\ \emph {et~al.}(2022)\citenamefont {Klein},
  \citenamefont {Pham}, \citenamefont {Thomsen}, \citenamefont {Curtis},
  \citenamefont {Denneulin}, \citenamefont {Lorke}, \citenamefont {Florian},
  \citenamefont {Steinhoff}, \citenamefont {Wiscons}, \citenamefont {Luxa}
  \emph {et~al.}}]{CrSBr_2}%
  \BibitemOpen
  \bibfield  {author} {\bibinfo {author} {\bibfnamefont {J.}~\bibnamefont
  {Klein}}, \bibinfo {author} {\bibfnamefont {T.}~\bibnamefont {Pham}},
  \bibinfo {author} {\bibfnamefont {J.}~\bibnamefont {Thomsen}}, \bibinfo
  {author} {\bibfnamefont {J.}~\bibnamefont {Curtis}}, \bibinfo {author}
  {\bibfnamefont {T.}~\bibnamefont {Denneulin}}, \bibinfo {author}
  {\bibfnamefont {M.}~\bibnamefont {Lorke}}, \bibinfo {author} {\bibfnamefont
  {M.}~\bibnamefont {Florian}}, \bibinfo {author} {\bibfnamefont
  {A.}~\bibnamefont {Steinhoff}}, \bibinfo {author} {\bibfnamefont
  {R.}~\bibnamefont {Wiscons}}, \bibinfo {author} {\bibfnamefont
  {J.}~\bibnamefont {Luxa}}, \emph {et~al.},\ }\bibfield  {title} {\bibinfo
  {title} {Control of structure and spin texture in the van der waals layered
  magnet crsbr},\ }\href@noop {} {\bibfield  {journal} {\bibinfo  {journal}
  {Nat. Commun.}\ }\textbf {\bibinfo {volume} {13}},\ \bibinfo {pages} {5420}
  (\bibinfo {year} {2022})}\BibitemShut {NoStop}%
\bibitem [{\citenamefont {Liu}\ \emph {et~al.}(2022)\citenamefont {Liu},
  \citenamefont {Li}, \citenamefont {Qiao}, \citenamefont {Wang}, \citenamefont
  {Zhang}, \citenamefont {Liu}, \citenamefont {Zhou}, \citenamefont {Shang},
  \citenamefont {Fang}, \citenamefont {Wang} \emph {et~al.}}]{twist_control1}%
  \BibitemOpen
  \bibfield  {author} {\bibinfo {author} {\bibfnamefont {C.}~\bibnamefont
  {Liu}}, \bibinfo {author} {\bibfnamefont {Z.}~\bibnamefont {Li}}, \bibinfo
  {author} {\bibfnamefont {R.}~\bibnamefont {Qiao}}, \bibinfo {author}
  {\bibfnamefont {Q.}~\bibnamefont {Wang}}, \bibinfo {author} {\bibfnamefont
  {Z.}~\bibnamefont {Zhang}}, \bibinfo {author} {\bibfnamefont
  {F.}~\bibnamefont {Liu}}, \bibinfo {author} {\bibfnamefont {Z.}~\bibnamefont
  {Zhou}}, \bibinfo {author} {\bibfnamefont {N.}~\bibnamefont {Shang}},
  \bibinfo {author} {\bibfnamefont {H.}~\bibnamefont {Fang}}, \bibinfo {author}
  {\bibfnamefont {M.}~\bibnamefont {Wang}}, \emph {et~al.},\ }\bibfield
  {title} {\bibinfo {title} {Designed growth of large bilayer graphene with
  arbitrary twist angles},\ }\href@noop {} {\bibfield  {journal} {\bibinfo
  {journal} {Nat. Mater.}\ }\textbf {\bibinfo {volume} {21}},\ \bibinfo {pages}
  {1263} (\bibinfo {year} {2022})}\BibitemShut {NoStop}%
\bibitem [{\citenamefont {Liao}\ \emph {et~al.}(2020)\citenamefont {Liao},
  \citenamefont {Wei}, \citenamefont {Du}, \citenamefont {Wang}, \citenamefont
  {Tang}, \citenamefont {Yu}, \citenamefont {Wu}, \citenamefont {Zhao},
  \citenamefont {Xu}, \citenamefont {Han} \emph {et~al.}}]{twist_control2}%
  \BibitemOpen
  \bibfield  {author} {\bibinfo {author} {\bibfnamefont {M.}~\bibnamefont
  {Liao}}, \bibinfo {author} {\bibfnamefont {Z.}~\bibnamefont {Wei}}, \bibinfo
  {author} {\bibfnamefont {L.}~\bibnamefont {Du}}, \bibinfo {author}
  {\bibfnamefont {Q.}~\bibnamefont {Wang}}, \bibinfo {author} {\bibfnamefont
  {J.}~\bibnamefont {Tang}}, \bibinfo {author} {\bibfnamefont {H.}~\bibnamefont
  {Yu}}, \bibinfo {author} {\bibfnamefont {F.}~\bibnamefont {Wu}}, \bibinfo
  {author} {\bibfnamefont {J.}~\bibnamefont {Zhao}}, \bibinfo {author}
  {\bibfnamefont {X.}~\bibnamefont {Xu}}, \bibinfo {author} {\bibfnamefont
  {B.}~\bibnamefont {Han}}, \emph {et~al.},\ }\bibfield  {title} {\bibinfo
  {title} {Precise control of the interlayer twist angle in large scale
  {MoS$_{2}$} homostructures},\ }\href@noop {} {\bibfield  {journal} {\bibinfo
  {journal} {Nat. Commun.}\ }\textbf {\bibinfo {volume} {11}},\ \bibinfo
  {pages} {2153} (\bibinfo {year} {2020})}\BibitemShut {NoStop}%
\bibitem [{\citenamefont {Pakdel}\ \emph {et~al.}(2024)\citenamefont {Pakdel},
  \citenamefont {Rasmussen}, \citenamefont {Taghizadeh}, \citenamefont {Kruse},
  \citenamefont {Olsen},\ and\ \citenamefont {Thygesen}}]{PES}%
  \BibitemOpen
  \bibfield  {author} {\bibinfo {author} {\bibfnamefont {S.}~\bibnamefont
  {Pakdel}}, \bibinfo {author} {\bibfnamefont {A.}~\bibnamefont {Rasmussen}},
  \bibinfo {author} {\bibfnamefont {A.}~\bibnamefont {Taghizadeh}}, \bibinfo
  {author} {\bibfnamefont {M.}~\bibnamefont {Kruse}}, \bibinfo {author}
  {\bibfnamefont {T.}~\bibnamefont {Olsen}},\ and\ \bibinfo {author}
  {\bibfnamefont {K.~S.}\ \bibnamefont {Thygesen}},\ }\bibfield  {title}
  {\bibinfo {title} {High-throughput computational stacking reveals emergent
  properties in natural van der waals bilayers},\ }\href@noop {} {\bibfield
  {journal} {\bibinfo  {journal} {Nat. Commun.}\ }\textbf {\bibinfo {volume}
  {15}},\ \bibinfo {pages} {932} (\bibinfo {year} {2024})}\BibitemShut
  {NoStop}%
\bibitem [{\citenamefont {Yasuda}\ \emph {et~al.}(2021)\citenamefont {Yasuda},
  \citenamefont {Wang}, \citenamefont {Watanabe}, \citenamefont {Taniguchi},\
  and\ \citenamefont {Jarillo-Herrero}}]{Stacking-exp1}%
  \BibitemOpen
  \bibfield  {author} {\bibinfo {author} {\bibfnamefont {K.}~\bibnamefont
  {Yasuda}}, \bibinfo {author} {\bibfnamefont {X.}~\bibnamefont {Wang}},
  \bibinfo {author} {\bibfnamefont {K.}~\bibnamefont {Watanabe}}, \bibinfo
  {author} {\bibfnamefont {T.}~\bibnamefont {Taniguchi}},\ and\ \bibinfo
  {author} {\bibfnamefont {P.}~\bibnamefont {Jarillo-Herrero}},\ }\bibfield
  {title} {\bibinfo {title} {Stacking-engineered ferroelectricity in bilayer
  boron nitride},\ }\href@noop {} {\bibfield  {journal} {\bibinfo  {journal}
  {Science}\ }\textbf {\bibinfo {volume} {372}},\ \bibinfo {pages} {1458}
  (\bibinfo {year} {2021})}\BibitemShut {NoStop}%
\bibitem [{\citenamefont {Vizner~Stern}\ \emph {et~al.}(2021)\citenamefont
  {Vizner~Stern}, \citenamefont {Waschitz}, \citenamefont {Cao}, \citenamefont
  {Nevo}, \citenamefont {Watanabe}, \citenamefont {Taniguchi}, \citenamefont
  {Sela}, \citenamefont {Urbakh}, \citenamefont {Hod},\ and\ \citenamefont
  {Ben~Shalom}}]{Stacking-exp2}%
  \BibitemOpen
  \bibfield  {author} {\bibinfo {author} {\bibfnamefont {M.}~\bibnamefont
  {Vizner~Stern}}, \bibinfo {author} {\bibfnamefont {Y.}~\bibnamefont
  {Waschitz}}, \bibinfo {author} {\bibfnamefont {W.}~\bibnamefont {Cao}},
  \bibinfo {author} {\bibfnamefont {I.}~\bibnamefont {Nevo}}, \bibinfo {author}
  {\bibfnamefont {K.}~\bibnamefont {Watanabe}}, \bibinfo {author}
  {\bibfnamefont {T.}~\bibnamefont {Taniguchi}}, \bibinfo {author}
  {\bibfnamefont {E.}~\bibnamefont {Sela}}, \bibinfo {author} {\bibfnamefont
  {M.}~\bibnamefont {Urbakh}}, \bibinfo {author} {\bibfnamefont
  {O.}~\bibnamefont {Hod}},\ and\ \bibinfo {author} {\bibfnamefont
  {M.}~\bibnamefont {Ben~Shalom}},\ }\bibfield  {title} {\bibinfo {title}
  {Interfacial ferroelectricity by van der waals sliding},\ }\href@noop {}
  {\bibfield  {journal} {\bibinfo  {journal} {Science}\ }\textbf {\bibinfo
  {volume} {372}},\ \bibinfo {pages} {1462} (\bibinfo {year}
  {2021})}\BibitemShut {NoStop}%
\bibitem [{\citenamefont {Rog{\'e}e}\ \emph {et~al.}(2022)\citenamefont
  {Rog{\'e}e}, \citenamefont {Wang}, \citenamefont {Zhang}, \citenamefont
  {Cai}, \citenamefont {Wang}, \citenamefont {Chhowalla}, \citenamefont {Ji},\
  and\ \citenamefont {Lau}}]{Stacking-exp3}%
  \BibitemOpen
  \bibfield  {author} {\bibinfo {author} {\bibfnamefont {L.}~\bibnamefont
  {Rog{\'e}e}}, \bibinfo {author} {\bibfnamefont {L.}~\bibnamefont {Wang}},
  \bibinfo {author} {\bibfnamefont {Y.}~\bibnamefont {Zhang}}, \bibinfo
  {author} {\bibfnamefont {S.}~\bibnamefont {Cai}}, \bibinfo {author}
  {\bibfnamefont {P.}~\bibnamefont {Wang}}, \bibinfo {author} {\bibfnamefont
  {M.}~\bibnamefont {Chhowalla}}, \bibinfo {author} {\bibfnamefont
  {W.}~\bibnamefont {Ji}},\ and\ \bibinfo {author} {\bibfnamefont {S.~P.}\
  \bibnamefont {Lau}},\ }\bibfield  {title} {\bibinfo {title} {Ferroelectricity
  in untwisted heterobilayers of transition metal dichalcogenides},\
  }\href@noop {} {\bibfield  {journal} {\bibinfo  {journal} {Science}\ }\textbf
  {\bibinfo {volume} {376}},\ \bibinfo {pages} {973} (\bibinfo {year}
  {2022})}\BibitemShut {NoStop}%
\bibitem [{\citenamefont {Sui}\ \emph {et~al.}(2024)\citenamefont {Sui},
  \citenamefont {Li}, \citenamefont {Qi}, \citenamefont {Jin}, \citenamefont
  {Lv}, \citenamefont {Wu}, \citenamefont {Liu}, \citenamefont {Zheng},
  \citenamefont {Liu}, \citenamefont {Ge} \emph {et~al.}}]{Stacking-exp4}%
  \BibitemOpen
  \bibfield  {author} {\bibinfo {author} {\bibfnamefont {F.}~\bibnamefont
  {Sui}}, \bibinfo {author} {\bibfnamefont {H.}~\bibnamefont {Li}}, \bibinfo
  {author} {\bibfnamefont {R.}~\bibnamefont {Qi}}, \bibinfo {author}
  {\bibfnamefont {M.}~\bibnamefont {Jin}}, \bibinfo {author} {\bibfnamefont
  {Z.}~\bibnamefont {Lv}}, \bibinfo {author} {\bibfnamefont {M.}~\bibnamefont
  {Wu}}, \bibinfo {author} {\bibfnamefont {X.}~\bibnamefont {Liu}}, \bibinfo
  {author} {\bibfnamefont {Y.}~\bibnamefont {Zheng}}, \bibinfo {author}
  {\bibfnamefont {B.}~\bibnamefont {Liu}}, \bibinfo {author} {\bibfnamefont
  {R.}~\bibnamefont {Ge}}, \emph {et~al.},\ }\bibfield  {title} {\bibinfo
  {title} {Atomic-level polarization reversal in sliding ferroelectric
  semiconductors},\ }\href@noop {} {\bibfield  {journal} {\bibinfo  {journal}
  {Nat. Commun.}\ }\textbf {\bibinfo {volume} {15}},\ \bibinfo {pages} {3799}
  (\bibinfo {year} {2024})}\BibitemShut {NoStop}%
\bibitem [{\citenamefont {Duerloo}\ \emph
  {et~al.}(2012{\natexlab{a}})\citenamefont {Duerloo}, \citenamefont {Ong},\
  and\ \citenamefont {Reed}}]{pie_BN}%
  \BibitemOpen
  \bibfield  {author} {\bibinfo {author} {\bibfnamefont {K.-A.~N.}\
  \bibnamefont {Duerloo}}, \bibinfo {author} {\bibfnamefont {M.~T.}\
  \bibnamefont {Ong}},\ and\ \bibinfo {author} {\bibfnamefont {E.~J.}\
  \bibnamefont {Reed}},\ }\bibfield  {title} {\bibinfo {title} {Intrinsic
  piezoelectricity in two-dimensional materials},\ }\href@noop {} {\bibfield
  {journal} {\bibinfo  {journal} {J. Phys. Chem. Lett.}\ }\textbf {\bibinfo
  {volume} {3}},\ \bibinfo {pages} {2871} (\bibinfo {year}
  {2012}{\natexlab{a}})}\BibitemShut {NoStop}%
\bibitem [{\citenamefont {Zhuang}\ \emph {et~al.}(2014)\citenamefont {Zhuang},
  \citenamefont {Johannes}, \citenamefont {Blonsky},\ and\ \citenamefont
  {Hennig}}]{pie_crs2}%
  \BibitemOpen
  \bibfield  {author} {\bibinfo {author} {\bibfnamefont {H.~L.}\ \bibnamefont
  {Zhuang}}, \bibinfo {author} {\bibfnamefont {M.~D.}\ \bibnamefont
  {Johannes}}, \bibinfo {author} {\bibfnamefont {M.~N.}\ \bibnamefont
  {Blonsky}},\ and\ \bibinfo {author} {\bibfnamefont {R.~G.}\ \bibnamefont
  {Hennig}},\ }\bibfield  {title} {\bibinfo {title} {Computational prediction
  and characterization of single-layer {CrS$_{2}$}},\ }\href@noop {} {\bibfield
   {journal} {\bibinfo  {journal} {Appl. Phys. Lett.}\ }\textbf {\bibinfo
  {volume} {104}} (\bibinfo {year} {2014})}\BibitemShut {NoStop}%
\bibitem [{\citenamefont {Ong}\ and\ \citenamefont
  {Reed}(2012)}]{pie_graphene}%
  \BibitemOpen
  \bibfield  {author} {\bibinfo {author} {\bibfnamefont {M.~T.}\ \bibnamefont
  {Ong}}\ and\ \bibinfo {author} {\bibfnamefont {E.~J.}\ \bibnamefont {Reed}},\
  }\bibfield  {title} {\bibinfo {title} {Engineered piezoelectricity in
  graphene},\ }\href@noop {} {\bibfield  {journal} {\bibinfo  {journal} {ACS
  Nano}\ }\textbf {\bibinfo {volume} {6}},\ \bibinfo {pages} {1387} (\bibinfo
  {year} {2012})}\BibitemShut {NoStop}%
\bibitem [{\citenamefont {Duerloo}\ \emph
  {et~al.}(2012{\natexlab{b}})\citenamefont {Duerloo}, \citenamefont {Ong},\
  and\ \citenamefont {Reed}}]{pie_tmdc}%
  \BibitemOpen
  \bibfield  {author} {\bibinfo {author} {\bibfnamefont {K.-A.~N.}\
  \bibnamefont {Duerloo}}, \bibinfo {author} {\bibfnamefont {M.~T.}\
  \bibnamefont {Ong}},\ and\ \bibinfo {author} {\bibfnamefont {E.~J.}\
  \bibnamefont {Reed}},\ }\bibfield  {title} {\bibinfo {title} {Intrinsic
  piezoelectricity in two-dimensional materials},\ }\href@noop {} {\bibfield
  {journal} {\bibinfo  {journal} {J. Phys. Chem. Lett.}\ }\textbf {\bibinfo
  {volume} {3}},\ \bibinfo {pages} {2871} (\bibinfo {year}
  {2012}{\natexlab{b}})}\BibitemShut {NoStop}%
\bibitem [{\citenamefont {Blonsky}\ \emph {et~al.}(2015)\citenamefont
  {Blonsky}, \citenamefont {Zhuang}, \citenamefont {Singh},\ and\ \citenamefont
  {Hennig}}]{pie_2d}%
  \BibitemOpen
  \bibfield  {author} {\bibinfo {author} {\bibfnamefont {M.~N.}\ \bibnamefont
  {Blonsky}}, \bibinfo {author} {\bibfnamefont {H.~L.}\ \bibnamefont {Zhuang}},
  \bibinfo {author} {\bibfnamefont {A.~K.}\ \bibnamefont {Singh}},\ and\
  \bibinfo {author} {\bibfnamefont {R.~G.}\ \bibnamefont {Hennig}},\ }\bibfield
   {title} {\bibinfo {title} {Ab initio prediction of piezoelectricity in
  two-dimensional materials},\ }\href@noop {} {\bibfield  {journal} {\bibinfo
  {journal} {ACS Nano}\ }\textbf {\bibinfo {volume} {9}},\ \bibinfo {pages}
  {9885} (\bibinfo {year} {2015})}\BibitemShut {NoStop}%
\bibitem [{\citenamefont {Lin}\ \emph {et~al.}(2022)\citenamefont {Lin},
  \citenamefont {Ponc{\'e}},\ and\ \citenamefont {Marzari}}]{imaginary1}%
  \BibitemOpen
  \bibfield  {author} {\bibinfo {author} {\bibfnamefont {C.}~\bibnamefont
  {Lin}}, \bibinfo {author} {\bibfnamefont {S.}~\bibnamefont {Ponc{\'e}}},\
  and\ \bibinfo {author} {\bibfnamefont {N.}~\bibnamefont {Marzari}},\
  }\bibfield  {title} {\bibinfo {title} {General invariance and equilibrium
  conditions for lattice dynamics in {1D}, {2D}, and {3D} materials},\
  }\href@noop {} {\bibfield  {journal} {\bibinfo  {journal} {npj Comput.
  Mater.}\ }\textbf {\bibinfo {volume} {8}},\ \bibinfo {pages} {236} (\bibinfo
  {year} {2022})}\BibitemShut {NoStop}%
\bibitem [{\citenamefont {Cong}\ \emph {et~al.}(2022)\citenamefont {Cong},
  \citenamefont {Liu}, \citenamefont {Xue}, \citenamefont {Liu}, \citenamefont
  {Liu},\ and\ \citenamefont {Shen}}]{imaginary2}%
  \BibitemOpen
  \bibfield  {author} {\bibinfo {author} {\bibfnamefont {A.}~\bibnamefont
  {Cong}}, \bibinfo {author} {\bibfnamefont {J.}~\bibnamefont {Liu}}, \bibinfo
  {author} {\bibfnamefont {W.}~\bibnamefont {Xue}}, \bibinfo {author}
  {\bibfnamefont {H.}~\bibnamefont {Liu}}, \bibinfo {author} {\bibfnamefont
  {Y.}~\bibnamefont {Liu}},\ and\ \bibinfo {author} {\bibfnamefont
  {K.}~\bibnamefont {Shen}},\ }\bibfield  {title} {\bibinfo {title}
  {Exchange-mediated magnon-phonon scattering in monolayer {CrI$_{3}$}},\
  }\href@noop {} {\bibfield  {journal} {\bibinfo  {journal} {Phys. Rev. B}\
  }\textbf {\bibinfo {volume} {106}},\ \bibinfo {pages} {214424} (\bibinfo
  {year} {2022})}\BibitemShut {NoStop}%
\bibitem [{\citenamefont {Cheng}\ and\ \citenamefont
  {Guan}(2021)}]{imaginary3}%
  \BibitemOpen
  \bibfield  {author} {\bibinfo {author} {\bibfnamefont {M.}~\bibnamefont
  {Cheng}}\ and\ \bibinfo {author} {\bibfnamefont {J.}~\bibnamefont {Guan}},\
  }\bibfield  {title} {\bibinfo {title} {Two-dimensional haeckelite ges with
  high carrier mobility and exotic polarization orders},\ }\href@noop {}
  {\bibfield  {journal} {\bibinfo  {journal} {Phys. Rev. Mater.}\ }\textbf
  {\bibinfo {volume} {5}},\ \bibinfo {pages} {054005} (\bibinfo {year}
  {2021})}\BibitemShut {NoStop}%
\end{thebibliography}%
\end{document}